\documentclass[showpacs,prb,twocolumn]{revtex4}
\usepackage{amssymb}
\usepackage{amsmath}
\usepackage{graphicx}
\usepackage{float}
\usepackage{amsfonts}
\begin{document}
\title{Josephson vortices and solitons inside pancake vortex lattice in layered superconductors}
\author{A. E. Koshelev}
\affiliation{Materials Science Division, Argonne National Laboratory, 9700 South Cass
Avenue, Argonne , IL 60439}
\date{\today}
\begin{abstract}
In very anisotropic layered superconductors a tilted magnetic
field generates crossing vortex lattices of pancake and Josephson
vortices (JVs). We study the properties of an isolated JV in the
lattice of pancake vortices.  JV induces deformations in the
pancake vortex crystal, which, in turn, substantially modify the
JV structure. The phase field of the JV is composed of two types
of phase deformations: the regular phase and vortex phase. The
phase deformations with smaller stiffness dominate. The
contribution from the vortex phase smoothly takes over with
increasing magnetic field. We find that the structure of the cores
experiences a smooth yet qualitative evolution with decrease of
the anisotropy. At large anisotropies pancakes have only small
deformations with respect to position of the ideal crystal while
at smaller anisotropies the pancake stacks in the central row
smoothly transfer between the neighboring lattice positions
forming a solitonlike structure. We also find that even at high
anisotropies pancake vortices strongly pin JVs and strongly
increase their viscous friction.
\end{abstract}
 \maketitle

\section{Introduction}

The vortex state in layered superconductors has a very rich phase
diagram in the multidimensional space of
temperature-field-anisotropy-field orientation. Especially
interesting subject is the vortex phases in layered
superconductors with very high anisotropy such as
Bi$_{2}$Sr$_{2}$CaCu$_{2}$O$_{x}$ (BSCCO). Relatively simple
vortex structures are formed when magnetic field is applied along
one of the principal axes of the layered structure. A magnetic
field applied perpendicular to the layers penetrates inside the
superconductor in the form of pancake vortices (PVs).
\cite{pancakes} PVs in different layers are coupled weakly via the
Josephson and magnetic interactions and form aligned stacks at low
fields and temperatures (PV stacks). These stacks are
disintegrated at the melting point. In another simple case of the
magnetic field applied parallel to the layers the vortex structure
is completely different. Such a field penetrates inside the
superconductor in the form of Josephson vortices
(JVs).\cite{BulJosLat,ClemJosLat,Kinkwalls} The JVs do not have
normal cores, but have rather wide nonlinear cores, of the order
of the Josephson length, located between two central layers. At
small in-plane fields JVs form the triangular lattice, strongly
stretched along the direction of the layers, so that JVs form
stacks aligned along the $c$ direction and separated by a large
distance in the in-plane direction.

A rich variety of vortex structures were theoretically predicted
for the case of tilted magnetic field, such as the kinked lattice,
\cite{BLK,Feinberg93,Kinkwalls}, tilted vortex chains
\cite{chains},  coexisting lattices with different
orientation.\cite{coex-lat} A very special situation exists in
highly anisotropic superconductors, in which the magnetic coupling
between the pancake vortices in different layers is stronger than
the Josephson coupling. In such superconductors a tilted magnetic
field creates a unique vortex state consisting of two
qualitatively different interpenetrating
sublattices.\cite{BLK,CrossLatLet}  This set of crossing lattices
(or combined lattice \cite{BLK}) contains a sublattice of
Josephson vortices generated by the component of the field
parallel to the layers, coexisting with a sublattice of stacks of
pancake vortices generated by the component of the field
perpendicular to the layers.
A basic reason for such an exotic ground state, as opposed to a
simple tilted vortex lattice, is that magnetic coupling energy is
minimal when pancake stacks are perfectly aligned along the $c$
axis. A homogeneous tilt of pancake lattice costs too much
magnetic coupling energy, while formation of Josephson vortices
only weakly disturbs the alignment of pancake stacks.

Even at high anisotropies JVs and pancake stacks have significant
attractive coupling.\cite{CrossLatLet} The strong mutual
interaction between the two sublattices leads to a very rich phase
diagram with many nontrivial lattice structures separated by phase
transitions.
At sufficiently small c-axis fields (10-50 gauss) a phase
separated state is formed: density of the pancake stacks located
at JVs becomes larger than the stack density outside
JVs.\cite{Huse,CrossLatLet} This leads to formation of dense stack
chains separated by regions of dilute triangular lattice in
between (mixed chain+lattice state). Such structures have been
observed in early decoration experiments \cite{Bolle91,Grig95}
and, more recently, by scanning Hall probe \cite{GrigNat01},
Lorentz microscopy \cite{MatsudaSci02}, and magnetooptical
technique.\cite{VlaskoPRB02,TokunagaPRB02} At very small c-axis
fields ($\sim$ several gauss) the regions of triangular lattice
vanish leaving only chains of stacks.\cite{GrigNat01} Moreover,
there are experimental indications \cite{GrigNat01} and
theoretical reasoning \cite{DodgsonPRB02} in favor of the phase
transition from the crossing configuration of pancake-stack chains
and JVs into chains of tilted vortices. JVs also modify the
interaction between pancake stacks leading to an attractive
interaction between the stacks at large
distances.\cite{BuzdinPRL02} As a consequence, one can expect
clustering of the pancake stacks at small concentrations.

Many unexpected observable effects can be naturally interpreted
within the crossing lattices picture. The underlying JV lattice
modifies the free energy of the vortex crystal state. An
observable consequence of this change is a shift of the melting
temperature. Strong support for the crossing-lattices ground state
is the linear dependence of the c-axis melting field on the
in-plane field observed within a finite range of in-plane
fields.\cite{OoiPRL99,CrossLatLet,Schmidt,KonczPhysC00,MirkovPRL01}
In an extended field range several melting regimes  have been
observed \cite{KonczPhysC00,MirkovPRL01} indicating several
distinct ground states of vortex matter in tilted fields.
Transitions between different ground state configurations have
also been detected by the features in the irreversible
magnetization.\cite{OoiPRB01,TokunagaPRB02A}
%

In this paper we consider in detail the properties of an isolated
JV in the pancake lattice. We mainly focus on the regime of a
dense pancake lattice, when many rows of the pancakes fit into the
JV core. The pancake lattice forms an effective medium for JVs
which determines their properties.  The dense pancake lattice
substantially modifies the JV structure. In general, the phase
field of the JV is built up from the continuous regular phase and
the phase perturbations created by pancake displacements. Such a
JV has a smaller core size and smaller energy as compared to the
ordinary JV.\cite{CrossLatLet} The pancake lattice also strongly
modifies the field and current distribution far away from the core
region.\cite{Savel01}

The key parameter which determines the structure of the JV core in
the dense pancake lattice is the ratio $\alpha=\lambda/\gamma s$,
where $\lambda$ is the in-plane London penetration depth, $\gamma$
is the anisotropy ratio, and $s$ is the period of layered
structure. The core structure experiences a smooth yet qualitative
evolution with decrease of this parameter. When $\alpha$ is small
(large anisotropies) pancakes have only small displacements with
respect to positions of the ideal crystal and the JV core occupies
several pancakes rows. In this situation the renormalization of
the JV core by the pancake vortices can be described in terms of
the continuous vortex phase which is characterized by its own
phase stiffness (effective phase stiffness
approach).\cite{CrossLatLet} At large $\alpha$ (small
anisotropies) the core shrinks to scales smaller than the distance
between pancake vortices. In this case pancake stacks in the
central row form soliton-like structure smoothly transferring
between the neighboring lattice position.

We consider dynamic properties of JVs in the case of small
$\alpha$: the critical pinning force which sticks the JV to the PV
lattice and the viscosity of the moving JV due to the traveling
displacement field in the PV lattice. The pinning force has a
nonmonotonic dependence on the c-axis magnetic field, $B_z$,
reaching maximum when roughly one pancake row fits inside the JV
core region. At higher fields the pinning force decays
exponentially $\propto \exp(-\sqrt{B_z/B_0})$. We study JV motion
through the PV lattice and find that the lattice strongly hinders
the mobility of JVs.

The paper is organized as follows. Section \ref{Sec:JVStructure}
is devoted to the static structure of an isolated JV inside the PV
lattice. In this section we
\begin{itemize}
\item consider small c-axis fields and calculate the crossing
energy of JV and PV stack (\ref{Sec:CrEn}); \item consider large
c-axis fields and introduce the ``effective phase stiffness''
approximation, which allows for simple description of JV structure
inside the dense PV lattice in the case of large anisotropy
(\ref{Sec:EffPhStiff}); \item investigate a large-scale behavior
and JV magnetic field (\ref{Sec:LargeScale}); \item analyze the JV
core quantitatively using numeric minimization of the total energy
and find crossover from the JV core structure to the soliton core
structure with decrease of anisotropy (\ref{Sec:QuantCore}); \item
formulate a simple model which describes the soliton core
structure for small anisotropies (\ref{Sec:soliton}).
\end{itemize} In Section \ref{Sec:JVPin} we consider pinning
of JV by the pancake lattice and calculate the field dependence of
the critical current at which the JV detaches from the PV lattice.
In Section \ref{Sec:JVVisc} we consider possible JV dynamic
regimes: dragging the pancake lattice by JVs and motion of JVs
through the pancake lattice. For the second case we calculate the
effective JV viscosity.

\section{\label{Sec:JVStructure}Structure and energy of Josephson vortex in pancake lattice}

\subsection{\label{Sec:GenRel}General relations}

Our calculations are based on the Lawrence-Doniach free energy in
the London approximation, which depends on the in-plane phases
$\phi_n(\mathbf{r})$ and vector-potential $\mathbf{A}(\mathbf{r})$
\begin{align}
F &  =\sum_{n}\int d^{2}\mathbf{r}\left[  \frac{J}{2}\left(
\mathbf{\nabla
}_{\perp}\phi_{n}-\frac{2\pi}{\Phi_{0}}\mathbf{A}_{\perp}\right)
^{2}\right.
\nonumber\\
&  \left.  +E_{J}\left(  1-\cos\left(  \phi_{n+1}-\phi_{n}-\frac{2\pi s}
{\Phi_{0}}A_{z}\right)  \right)  \right]  \nonumber\\
&  +\int d^{3}\mathbf{r}\frac{\mathbf{B}^{2}}{8\pi},\label{LDen}
\end{align}
where
\begin{equation}
J\equiv\frac{s\Phi_{0}^{2}}{\pi\left(  4\pi\lambda\right)  ^{2}}\equiv
\frac{s\varepsilon_{0}}{\pi}\;\text{and }E_{J}\equiv\frac{\Phi_{0}^{2}}
{s\pi\left(  4\pi\lambda_{c}\right)  ^{2}}
\end{equation}
are the phase stiffness and the Josephson coupling energy,
$\lambda \equiv\lambda_{ab}$ and $\lambda_{c}$ are the components
of the London penetration depth and $s$ is the interlayer
periodicity. We use the London gauge, $\mathrm{div}\mathbf{A}=0$.
We assume that the average magnetic induction $\mathbf{B}$ inside
the superconductor is fixed.\cite{BvsH} The $c$ component of the
field fixes the concentration of the pancake vortices $n_{v}\equiv
B_z/\Phi_0$ inside one layer. The in-plane phases $\phi_{n}$ have
singularities at the positions of pancake vortices
$\mathbf{R}_{in}$ inside the layers,
\[
\left [\mathbf{\nabla}\times \mathbf{\nabla}\phi_{n}\right
]_z=2\pi\sum_{i}\delta\left( \mathbf{r} -\mathbf{R}_{in}\right)  .
\]
The major obstacle, preventing a full analytical consideration of
the problem, is the nonlinearity coming from the Josephson term. A
useful approach for superconductors with weak Josephson coupling
is to split the phase and vector-potential into the vortex and
regular contributions, $\phi_{n}=\phi_{vn} +\phi_{rn}$ and
$\mathbf{A}=\mathbf{A}_{v}+\mathbf{A}_{r}$. The vortex
contributions minimize the energy for fixed positions of pancake
vortices at $E_{J}=0$ and give magnetic interaction energy for the
pancake vortices. One can express this part of energy via the
vortex coordinates $\mathbf{R}_{n,i}$. In general, the regular
contributions may include phases and vector-potentials of the
Josephson vortices. The total energy naturally splits into the
regular part $F_{r}$, the energy of magnetic interactions between
pancakes $F_{M}$, and the Josephson energy $F_{J}$, which couples
the regular and vortex degrees of freedom,
\begin{equation}
F=F_{r}+F_{M}+F_{J} \label{splitLDen}
\end{equation}
with
\begin{widetext}
\begin{align}
F_{r}\left[  \phi_{rn},\mathbf{A}_{r}\right]   &  =\sum_{n}\int d^{2}
\mathbf{r}\frac{J}{2}\left(  \mathbf{\nabla}\phi_{rn}-\frac{2\pi}{\Phi_{0}
}\mathbf{A}_{r\perp}\right)  ^{2}+\int d^{3}\mathbf{r}\frac{\mathbf{B}_{r}
^{2}}{8\pi},\label{Fr}\\
F_{M}\left[  \mathbf{R}_{n,i}\right]   &  =\frac{1}{2}\sum_{n,m,i,j}
U_{M}(\mathbf{R}_{n,i}-\mathbf{R}_{m,j},n-m),\label{Fv}\\
F_{J}\left[  \phi_{rn},\mathbf{A}_{r},\mathbf{R}_{n,i}\right]   &  =\sum
_{n}\int d^{2}\mathbf{r}E_{J}\left(  1-\cos\left(  \phi_{n+1}-\phi_{n}
-\frac{2\pi s}{\Phi_{0}}A_{z}\right)  \right)  , \label{FJ}
\end{align}
and
\begin{align}
U_{M}(\mathbf{R},n)  &  =\frac{J}{2\pi}\int d\mathbf{k}\int_{-\pi}^{\pi
}dq\frac{\exp[i\mathbf{kR}+iqn)]}{k^{2}[1+\lambda^{-2}(k^{2}+2(1-\cos
q)/s^{2})^{-1}]}\nonumber\\
&  \approx2\pi J\left[  \ln\frac{L}{R}\left[  \delta_{n}-\frac{s}{2\lambda
}\exp\left(  -\frac{s|n|}{\lambda}\right)  \right]  +\frac{s}{4\lambda
}u\left(  \frac{r}{\lambda},\frac{s|n|}{\lambda}\right)  \right]
\label{MagInter}
\end{align}
\end{widetext}
is the magnetic interaction between pancakes \cite{pancakes} where
\begin{equation}
u\left(  r,z\right)  \equiv\exp(-z)E_{1}\left(  r-z\right)  +\exp
(z)E_{1}\left(  r+z\right)  , \label{urz}
\end{equation}
$E_{1}(u)=\int_{u}^{\infty}\left(  \exp(-v)/v\right)  dv$ is the
integral exponent ($E_{1}(u)\approx-\gamma_E-\ln u+u\ $at$\;u\ll1$
with $\gamma_E \approx 0.577$), $r\equiv\sqrt{R^{2}+(ns)^{2}}$,
and $L$ is a cutoff length.

In this paper we focus on the crystal state. If the pancake coordinates have
only small deviations from the positions $\mathbf{R}_{i}^{(0)}$ of the ideal
triangular lattice, $\mathbf{R}_{ni}=\mathbf{R}_{i}^{(0)}+\mathbf{u}_{ni}$
then $F_{v}$ reduces to the energy of an ideal crystal $F_{cr}$ plus the
magnetic elastic energy $F_{M-el}$ consisting of the shear and compression
parts,
\begin{equation}
F_{M-el}\!=\!\int\!\frac{d^{3}\mathbf{k}}{\left(\!  2\pi\!\right)
^{3}}\left[ \frac{U_{t}(\mathbf{k})}{2}\left\vert
u_{t}(\mathbf{k})\right\vert ^{2}
+\frac{U_{l}(\mathbf{k})}{2}\left\vert
u_{l}(\mathbf{k})\right\vert
^{2}\right]  , \label{FMel}
\end{equation}
where
\[
\mathbf{u}_{ni}\!\equiv\!\int\!\frac{d^{3}\mathbf{k}}{(2\pi)^{3}}\exp\left(
i\mathbf{k}_{\perp}\mathbf{R}_{i}^{(0)}\!+\!ik_{z}sn\right) \left[
\mathbf{e}_{t}u_{t}(\mathbf{k})\!+\!\mathbf{e}_{l}u_{l}(\mathbf{k})\right]
\]
($\mathbf{e}_{l}$ ($\mathbf{e}_{t}$) is the unit vector parallel
(orthogonal) to $\mathbf{k}_{\perp}$),
\begin{align*}
U_{t}(\mathbf{k})  &  =C_{66}k_{\perp}^{2}+U_{44}(\mathbf{k}),\\
U_{l}(\mathbf{k})  &  =U_{11}(\mathbf{k})+U_{44}(\mathbf{k}),
\end{align*}
$U_{11}$ is the compression stiffness, $C_{66}$ is the shear
modulus. In particular, at high fields,
$B_{z}\gg\Phi_{0}/\lambda^{2}$, we have
\begin{align*}
U_{11}&\equiv
C_{11}(\mathbf{k})k_{\perp}^{2}\approx\frac{B_{z}^{2}}
{4\pi\lambda^{2}}\left(  1-\frac{k_{\perp}^{2}}{16\pi
n_{v}}\right) ,\\
C_{66}&=n_{v}\epsilon_{0}/4
\end{align*}
The magnetic tilt stiffness,\cite{KoshKes93} $U_{44}(\mathbf{k})$,
is given by interpolation formula, which takes into account
softening due to pancake fluctuations,
\begin{align}
U_{44}(\mathbf{k})&\equiv C_{44}(k_{z})k_{z}^{2}\nonumber\\
&=\frac{B_{z}\Phi_{0}}
{2(4\pi)^{2}\lambda^{4}}\ln\left(  1+\frac{r_{cut}^{2}}{k_{z}^{-2}+r_{w}^{2}
}\right)  \label{TiltStiff}
\end{align}
with $r_{w}^{2}=\left\langle (u_{n+1}-u_{n})^{2}\right\rangle $,
$r_{cut}\approx\lambda$ at $a>\lambda$ and $r_{cut}\approx a/4.5$
at $a<\lambda$,\cite{TiltSCHA} with $a=\sqrt{2/(\sqrt{3}n_v)}$
being the lattice constant. At finite Josephson energy
minimization of the energy with respect to the phases at fixed
pancake positions leads to the Josephson term in the tilt
stiffness energy.\cite{Horovitz}

The major focus of this paper is the structure of JV core. This
requires analysis of the pancake displacements and regular phase
at distances $r_{\perp}\ll\lambda_{c}$ and $z\ll a,\lambda$ from
the vortex center. At these distances the main contribution to the
energy is coming from the kinetic energy of supercurrents and
magnetic screening can be neglected. The structure of energy is
significantly simplified: one can neglect the field contributions
in the regular and Josephson energy terms, i.e., drop
$\mathbf{A}_{r}$:
\begin{widetext}
\begin{align}
F_{r}\left[  \phi_{rn},\mathbf{A}_{r}\right]  \!  &  \rightarrow\!
F_{r}\left[  \phi_{rn}\right]  =\sum_{n}\int d^{2}\mathbf{r}\frac{J}{2}\left(
\mathbf{\nabla}\phi_{rn}\right)  ^{2},\\
F_{J}\left[  \phi_{rn},\mathbf{A}_{r},\mathbf{R}_{n,i}\right]  \!  &
\rightarrow\! F_{J}\left[  \phi_{rn},\mathbf{R}_{n,i}\right]  =\sum_{n}\int
d^{2}\mathbf{r}E_{J}\left(  1-\cos\left(  \phi_{n+1}-\phi_{n}\right)  \right)
,
\end{align}
\end{widetext}
and use asymptotics $r_{cut}k_{z}\gg1$ in the tilt stiffness
(\ref{TiltStiff}). Behavior at large distances
$r_{\perp}\sim\lambda_{c}$ and $z\sim a,\lambda$ is important for
accurate evaluation of the cutoff in the logarithmically diverging
energy of the Josephson vortex. In this range the Josephson term
can be linearized and one can use the anisotropic London
theory.\cite{Savel01}

\subsection{\label{Sec:CrEn}Small $c$ axis field: Crossing energy}

\begin{figure}
[ptb]
\begin{center}
\includegraphics[clip,width=3.2in ] {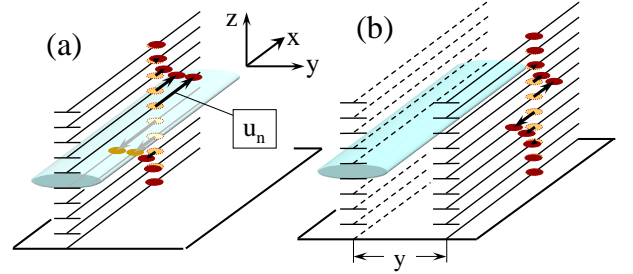}
\caption{Configuration of the
pancake stack crossing the Josephson vortex:(a) the stack located
in the center of JV core and (b) the stack located at a finite
distance $y$ from JV center.} \label{Fig-CrossConf}
\end{center}
\end{figure}

At small fields and high anisotropy factor pancake vortices do not
influence much structure of JVs. However, there is a finite
interaction energy between pancake stack and JV (crossing energy)
which causes spectacular observable effects, including formation
of the mixed chain-lattice
state.\cite{Bolle91,Grig95,MatsudaSci02,Huse,CrossLatLet} We
consider a JV located between the layers $0$ and $1$ and directed
along $x$ axis with center at $y=0$ and a pancake stack located at
$x=0$ and at distance $y$ from the JV center. We will calculate
structure of the pancake stack and the crossing energy. The JV
core structure is defined by the phases $\phi_{n}(y)$ obeying the
following equation
\begin{equation}
\frac{d^2\phi_{n}}{d\tilde{y}^2}+\sin(\phi_{n+1}-\phi_{n})-\sin(\phi_{n}-\phi_{n-1})=0
\label{JVphaseEq}
\end{equation}
with $\tilde{y}=y/\lambda_{J0}$.  An accurate numerical solution
of this equation has been obtained in Ref.\
\onlinecite{Kinkwalls}. It is described by the approximate
interpolation formula\cite{PhaseDistNote}
\begin{align}
\phi_{n}(\tilde{y})  &  \approx\arctan\frac{n-1/2}{\tilde{y}}+\frac
{0.35(n-1/2)\tilde{y}}{\left(  (n-1/2)^{2}+\tilde{y}^{2}+0.38\right)  ^{2}
}\nonumber\\
&  +\frac{8.81(n-1/2)\tilde{y}\left(
\tilde{y}^{2}-(n-1/2)^{2}+2.77\right) }{\left(
(n-1/2)^{2}+\tilde{y}^{2}+2.02\right) ^{4}}\label{JVPhase}
\end{align}
Interaction between the pancake stack and JV appears due to the
pancake displacements $u_{n}$ under the action of the JV in-plane
currents $j_{n}(y)$ (see Fig.\ \ref{Fig-CrossConf}). In the regime
of very weak interlayer coupling the energy of the deformed
pancake stack is given by
\begin{equation}
E_{\times}(y)=\int\frac{dk_{z}}{2\pi}\frac{U_{M}(k_{z})}{2}\left\vert
u(k_{z})\right\vert ^{2}-\sum_{n}\frac{s\Phi_{0}}{c}j_{n}(y)u_{n}
\label{Ecross-kz}
\end{equation}
where
$U_{M}(k_{z})=\frac{\Phi_{0}^{2}}{2(4\pi)^{2}\lambda^{4}}\ln\left(
1+\frac{\lambda^{2}}{k_{z}^{-2}+r_{w}^{2}}\right)  $ is the
magnetic tilt stiffness of the pancake stack,
\[
j_{n}(y)\approx\frac{2c\Phi_{0}}{\left(  4\pi\lambda\right)  ^{2}\gamma
s}p_{n}\left(  \frac{y}{\gamma s}\right)
\]
with $p_{n}(\tilde{y})\equiv d\phi_{n}(\tilde{y})/d\tilde{y}$
being the reduced superfluid momentum, and the JV phase
$\phi_{n}(\tilde{y})$ is given by approximate formula
(\ref{JVPhase}). In particular, $p_{n}(0)=-C_{n}/(n-1/2)$ with
$C_{n}\rightarrow1$ at large $n$. Using the precise numerical
phases $\phi_{n}(\tilde{y})$ we obtain the interpolation formula
\[
C_{n}\approx1-0.265/((n-0.835)^{2}+0.566),
\]
giving $C_{1}\approx0.55$ and $C_{2}\approx0.86$. At large
distances from the core, $n,\tilde{y}\gg1$, $p_{n}(\tilde{y})$ is
given by
\[
p_{n}(\tilde{y})=-\frac{n-1/2}{(n-1/2)^{2}+\tilde{y}^{2}}.
\]

Displacements in the core region have typical wave vectors
$k_{z}\sim\pi/s$. In this range one can neglect $k_{z}$-dependence
of $U_{M}(k_{z})$,
\[U_{M}\approx\frac{\Phi_{0}^{2}}{(4\pi)^{2}\lambda^{4}}\ln\frac{\lambda}
{r_{w}},\]
and rewrite Eq.\ (\ref{Ecross-kz}) as
\begin{equation}
E_{\times}(y)=\sum_{n}\left(  \frac{sU_{M}}{2}u_{n}^{2}-\frac{s\Phi_{0}}
{c}j_{n}(y)u_{n}\right)  . \label{Ecross-n}
\end{equation}
We neglected weak logarithmic dependence of the tilt stiffness on
displacements and the parameter $r_w$ in $U_{M}$ is just a typical
value of $|u_{n+1}-u_{n}|$. Minimizing this energy with respect to
$u_{n}$, we obtain the pancake displacements
\[
u_{n}(y)=\frac{\Phi_{0}}{c}\frac{j_{n}(y)}{U_{M}}\approx\frac{2\lambda^{2}
}{\gamma s\ln\left(  \lambda/u_{n}(y)\right)  }p_{n}\left[
\frac{y}{\gamma s}\right]
\]
with $v_{n}(y)\approx\left\vert u_{n}(y)-u_{n-1}(y)\right\vert $ and crossing
energy at finite distance $y$ between the crossing point and the center of
JV\ core
\begin{align}
E_{\times}(y)  &  =-\frac{s}{2U_{M}}\sum_{n=-\infty}^{\infty}\left(
\frac{\Phi_{0}j_{n}(y)}{c}\right)  ^{2}\nonumber\\
& \approx-\frac{\Phi_{0}^{2}}{4\pi^{2}\gamma^{2}s\ln(\lambda/u_{1}
(y))}A_{\times}\left(  \frac{y}{\gamma s}\right) \label{CrossEn-y}
\end{align}
with
\[
A_{\times}(\tilde{y})=\sum_{n=1}^{\infty}\left[ p_{n}\left( \tilde
{y}\right)  \right]  ^{2}\\
=\sum_{n=-\infty}^{\infty}\left( 1-\cos(\phi_{n+1}-\phi_n)
\right),
\]
where the second identity can be derived from Eq.\
(\ref{JVphaseEq}).  In particular, for the pancake stack located
at the JV\ center, $A_{\times}(0)= 2 $ (exact value) and
\begin{equation}
E_{\times}(0)\approx-\frac{\Phi_{0}^{2}}{2\pi^{2}\gamma^{2}s\ln(3.5\gamma
s/\lambda)}. \label{CroosEn-0}
\end{equation}
The maximum pancake displacement in the core is given by
\begin{equation}
u_{1}(0)\approx\frac{2.2\lambda^{2}}{\gamma s\ln\left(  2\gamma
s/\lambda\right)  }. \label{Maxu-smallB}
\end{equation}
At large distances, $y\gg\gamma s$, using asymptotics
$A_{\times}(\tilde {y})\approx\pi/4\tilde{y}$, we obtain
\begin{equation}
E_{\times}(y)\approx-\frac{\Phi_{0}^{2}}{16\pi\gamma y\ln\left(  \frac{\beta
y^{2}}{\lambda\gamma s}\right)  },\;\text{with }\beta\sim1.
\label{CroosEn-Large}
\end{equation}
Using numerical calculations, we also obtain the following approximate
interpolation formula for the function $A_{\times}(\tilde{y})$, valid for the
whole range of $\tilde{y}$,
\begin{align}
A_{\times}(\tilde{y})&\approx\frac{\pi/4}{\sqrt{\tilde{y}^{2}+y_{0}^{2}}
}\left(  1+\frac{a}{\tilde{y}^{2}+y_{0}^{2}}\right)
\label{AcrosInt}\\
y_{0}&\approx 0.93,\;a\approx1.23\nonumber
\end{align}
Eqs.\ (\ref{CrossEn-y}) and (\ref{AcrosInt}) determine the
crossing energy at finite distance $y$ between the crossing point
and the center of the JV core. Below we will use this result to
calculate the pinning force which binds the JV to the dilute
pancake lattice.

\subsection{\label{Sec:EffPhStiff}Large $c$ axis field. Approximation of the effective phase
stiffness}

At high $c$ axis fields pancakes substantially modify the JV
structure. Precise analysis of the JV core in the pancake lattice
for the general case requires tedious consideration of many energy
contributions (see Section \ref{Sec:QuantCore} below). The
situation simplifies considerably in the regime of very high
anisotropy $\gamma\gg\lambda/s$. In this case one can conveniently
describe the JV structure in terms of the effective phase
stiffness, which allows us to reduce the problem of a JV in the
pancake lattice to the problem of an ordinary JV at $B_{z}=0$. In
Ref.\ \onlinecite{CrossLatLet} this approach has been used to
derive JV\ structures at high fields
$B_{z}\gg\Phi_{0}/4\pi\lambda^{2}$. The approach is based on the
observation that smooth transverse lattice deformations
$\mathbf{u}_{tn}(\mathbf{r})$ produce large-scale phase variations
$\phi _{vn}(\mathbf{r})$ with $\mathbf{\nabla}\phi_{vn}=2\pi
n_{v}\mathbf{e} _{z}\times\mathbf{u}_{tn}$. This allows us to
express the transverse part of the elastic energy, $F_{v-t}$, in
terms of $\phi_{vn}(\mathbf{r})$:
\begin{equation}
F_{v-t}=\int\frac{d\mathbf{k}}{(2\pi)^{3}}\frac{J_{v}(B_{z},\mathbf{k})}
{2s}k_{\perp}^{2}\left|  \phi_{v}(\mathbf{k})\right|  ^{2},
\label{ShearEnPhase}
\end{equation}
with the effective phase stiffness $J_{v}(B_{z},\mathbf{k})$,
\begin{align}
J_{v}(B_{z},\mathbf{k})  &  =\frac{s\left(  C_{66}k_{\perp}^{2}+U_{44}\right)
}{\left(  2\pi n_{v}\right)  ^{2}}\label{PhStiff-k}\\
&  \approx\frac{J}{8\pi n_{v}\lambda^{2}}\left(  \frac{\lambda^{2}k_{\perp
}^{2}}{2}+\ln\left(  1+\frac{r_{cut}^{2}}{k_{z}^{-2}+r_{w}^{2}}\right)
\right)  .\nonumber
\end{align}
Replacing the discrete lattice displacements by the smooth phase
distribution is justified at fields $B_{z}>\Phi_{0}/(\gamma
s)^{2}$. The structure of the JV core is determined by phase
deformations with the typical wave vectors $k_{\perp }\sim1/\gamma
s<2/\lambda$ and $k_{z}\sim\pi/s>1/r_{w}$. In this range the
vortex phase stiffness is $\mathbf{k}$-independent, similar to the
usual phase stiffness,
\begin{equation}
J_{v}(B_{z})\approx J\frac{B_{\lambda}}{B_{z}},\;\;B_{\lambda}\equiv\frac
{\Phi_{0}}{4\pi\lambda^{2}}\ln\frac{r_{cut}}{r_{w}}. \label{VorPhStiff}
\end{equation}
The phase stiffness energy (\ref{ShearEnPhase}) has to be
supplemented by the Josephson energy. In the core region we can
neglect the vector-potentials and write the total energy in terms
of $\phi_{rn}$ and $\phi_{v}$ as
\begin{widetext}
\begin{equation}
F=\sum_{n}\int d^{2}\mathbf{r}\left(  \frac{J}{2}\left(
\nabla\phi _{rn}\right)  ^{2}+\frac{J_{v}}{2}\left(
\nabla\phi_{vn}\right)  ^{2} +E_{J}\left(  1-\cos\left(
\phi_{n+1}-\phi_{n}\right)  \right)  \right)  ,
\label{PhaseEn}
\end{equation}
\end{widetext}
where, again, $\phi_{n}\equiv\phi_{rn}+\phi_{vn}$ is the total phase.

We now investigate the core structure on the basis of the energy
(\ref{PhaseEn}). Eliminating the regular phase,
$\phi_{rn}=\phi_{n}-\phi_{vn}$, and varying the energy with
respect to $\phi_{vn}$ at fixed $\phi_{n}$, we obtain the equation
\[
J\Delta\phi_{n}-\left(  J_{v}+J\right)  \Delta\phi_{vn}=0,
\]
which gives
\begin{equation}
\phi_{vn}=\frac{J}{J_{v}+J}\phi_{n}. \label{VortPhase}
\end{equation}
Substituting this relation back into the energy (\ref{PhaseEn}),
we express it in terms of the total phase
\begin{equation}
F\!=\!\sum_{n}\int\! d^{2}\mathbf{r}\!\left(\!
\frac{J_{\mathrm{eff}}}{2}\left( \nabla\phi_{n}\right)
^{2}+E_{J}\left(  1-\cos\left(  \phi_{n+1}-\phi
_{n}\right)  \right)  \right)  \label{EffEn}
\end{equation}
with the effective phase stiffness $J_{\mathrm{eff}}$,
\begin{equation}
J_{\mathrm{eff}}^{-1}=J^{-1}+J_{v}^{-1}\;\text{or }J_{\mathrm{eff}}=\frac
{J}{1+B_{z}/B_{\lambda}}. \label{Jeff}
\end{equation}
Note that the smallest stiffness from $J$ and $J_{v}$ dominates in
$J_{\mathrm{eff}}$.\cite{noteDodgsonPRL99} From Eq.\ (\ref{EffEn})
we obtain equation for the equilibrium phase
\begin{equation}
J_{\mathrm{eff}}\nabla_{y}^{2}\phi_{n}+E_{J}\left[  \sin\left(  \phi
_{n+1}-\phi_{n}\right)  -\sin\left(  \phi_{n}-\phi_{n-1}\right)  \right]  =0,
\label{PhaseEq}
\end{equation}
which has the same form as at zero $c$ axis field, except that the
bare phase stiffness is replaced with the effective phase
stiffness $J_{\mathrm{eff}}$. For the Josephson vortex located
between the layers $0$ and $1$ the phase satisfies the conditions
\begin{equation}
\phi_{1}-\phi_{0}\rightarrow
\genfrac{\{}{.}{0pt}{}{0,\;y\rightarrow\infty}{2\pi,\;y\rightarrow-\infty}
\label{JVcond}
\end{equation}
Far away from the nonlinear core the phase has the usual form for the vortex
in anisotropic superconductor
\begin{equation}
\phi_{n}(y)\approx\arctan\frac{\lambda_{J}(n-1/2)}{y} \label{PhaseDist}
\end{equation}
where the effective Josephson length
\[
\lambda_{J}=\sqrt{J_{\mathrm{eff}}
/E_{J}}=\lambda_{J0}/\sqrt{1+B_{z}/B_{\lambda}}
\]
determines the size of the nonlinear core. Therefore, at low
temperatures the JV core shrinks in the presence of the $c$ axis
magnetic field  due to softening of the in-plane phase
deformations. A number of pancake rows within the JV\ core can be
estimated as
\begin{equation}
N_{\mathrm{rows}}\approx\frac{\lambda_{J0}}{\lambda}\sqrt{\frac{\ln
(r_{\mathrm{cut}}/r_{w})}{4\pi}}\sqrt{\frac{B_{z}}{B_{\lambda}+B_{z}}}.
\label{Nrows}
\end{equation}
At $B_{z}>B_{\lambda}$ it is almost independent on the field. An approximate
solution of Eq.\ (\ref{PhaseEq}) is given by Eq.\ (\ref{JVPhase}), where the
bare Josephson length $\lambda_{J0}$ has to be replaced by the renormalized
length $\lambda_{J}$.

The JV energy per unit length, $\mathcal{E}_{JV}$, is given by
\begin{equation}
\mathcal{E}_{JV}=\pi\sqrt{E_{J}J_{\mathrm{eff}}}\ln\frac{L}{s}, \label{EnJV}
\end{equation}
where $L$ is the cutoff length, which is determined by screening
at large distances and will be considered below, in Section
\ref{Sec:LargeScale}. From Eqs.\ (\ref{VortPhase}) and
(\ref{Jeff}) we obtain that the partial contribution of the vortex
phase in the total phase,
\begin{equation}
\phi_{vn}=\frac{B_{z}}{B_{z}+B_{\lambda}}\phi_{n}, \label{VortPhaseFraction}
\end{equation}
continuously grows from $0$ at $B_{z}\ll B_{\lambda}$ to $1$ at
$B_{z}\gg B_{\lambda}$. From the last equation one can estimate
pancake displacements
\begin{align}
u_{x,n}(y)  &  =\frac{\Phi_{0}/2\pi}{B_{z}+B_{\lambda}}\nabla_{y}\phi
_{n}\nonumber\\
&  \approx-\frac{\Phi_{0}/2\pi}{B_{z}+B_{\lambda}}\frac{\lambda_{J}
(n-1/2)}{y^{2}+\left(  \lambda_{J}(n-1/2)\right)  ^{2}}. \label{CoreDispl}
\end{align}
The maximum displacement in the core can be estimated as
\begin{equation}
u_{x,0}(0)\approx\frac{2.2\lambda^{2}}{\lambda_{J0}\ln(r_{cut}/r_{w}
)\sqrt{1+B_{z}/B_{\lambda}}}. \label{MaxDispl}
\end{equation}
At $B_{z}\gg B_{\lambda}$ this equation can be rewritten in the form
\begin{equation}
\frac{u_{x,0}(0)}{a}\approx\frac{0.58\lambda}{\lambda_{J0}\sqrt{\ln(a/r_{w})}
}, \label{MaxDisplHighB}
\end{equation}
which shows that condition for applicability of the linear
elasticity, $u_{x,0}(0)\lesssim0.2a$, is satisfied if
$\gamma\gtrsim3\lambda/s$. Eqs.\ (\ref{PhaseDist}) and
(\ref{EnJV}) describe smooth evolution of the JV structure with
increase of concentration of pancakes starting from the usual
vortex at $B_{z}=0$. It is quantitatively valid only at very high
anisotropies $\gamma\gg\lambda/s$ and at low temperatures, when
one can neglect fluctuation suppression of the Josephson energy.
Thermal motion of the PVs at finite temperatures induces the
fluctuating phase $\tilde{\phi}_{n,n+1}$\ and suppresses the
effective Josephson energy, $E_{J}\rightarrow CE_{J}$\ where
$C\equiv\left\langle \cos\tilde{\phi}_{n,n+1}\right\rangle $. This
leads to reduction of the JV energy and thermal expansion of its
core.

In the range $\Phi_{0}/(\gamma s)^{2}<B_{z}<B_{\lambda}$ the
``crossing energy'' regime of Section \ref{Sec:CrEn} overlaps with
the applicability range of the effective phase stiffness
approximation. To check the consistency of these approximations we
calculate correction to the JV energy at small fields summing up
the crossing energies and compare the result with prediction of
the ``effective phase stiffness'' approximation. The correction to
the JV energy is given by
\begin{align}
\delta\mathcal{E}_{JV}  &  =\frac{1}{a}\sum_{m=-M}^{M}E_{\times}
(mb)\nonumber\\
&  \approx- \frac{\Phi_{0}B_{z}}{8\pi\gamma\ln(\lambda/u_{1}(0))}\ln
\frac{L_{y}}{\gamma s}, \label{dJVen}
\end{align}
where $L_{y}=Mb$ is the long-range cutoff length. On the other hand,
Eqs.\ (\ref{Jeff}) and (\ref{EnJV}) give at $B_{z}\ll B_{\lambda}$
\begin{equation}
\delta\mathcal{E}_{JV}=\pi\sqrt{E_{J}J}\frac{B_{z}}{2B_{\lambda}}\ln\frac
{L}{s}.
\end{equation}
This result is identical to Eq.\ (\ref{dJVen}) except for
expressions under the logarithms, which are approximate in both
cases.

\subsection{\label{Sec:LargeScale}Large-scale behavior. Screening lengths}

In this Section we consider the JV structure at large distances
from the core, $n\gg1,\;y\gg\lambda_{J}$. At large distance
screening of supercurrents becomes important and one can not
neglect the vector-potential any more. At these scales the phase
changes slowly from layer to layer so that one can expand the
Josephson energy in Eq.\ (\ref{FJ}) with respect to the phase
difference and use the continuous approximation,
$\phi_{n+1}-\phi_{n}-\frac{2\pi s}{\Phi_{0} }A_{z}\rightarrow
s\left( \mathbf{\nabla}_{z}\phi-\frac{2\pi}{\Phi_{0}} A_{z}\right)
$,
\begin{equation}
F_{J}\left[  \phi_{n},\mathbf{A}\right]  \rightarrow\tilde{F}_{J}\left[
\phi,\mathbf{A}\right]  =\int d^{3}\mathbf{r}\frac{sE_{J}}{2}\left(
\mathbf{\nabla}_{z}\phi-\frac{2\pi}{\Phi_{0}}A_{z}\right)  ^{2}. \label{FJLon}
\end{equation}
This reduces the Lawrence-Doniach model defined by Eqs.\
(\ref{splitLDen}), (\ref{Fr}), (\ref{Fv}), and (\ref{FJ}) to the
anisotropic London model. Within this model the JV structure
outside the core region has been investigated in detail by
Savel'ev {\emph et. al.} \cite{Savel01}. In this section we
reproduce JV structure at large distances using the effective
phase stiffness approach. For the vortex energy one still can use
Eq.\ (\ref{ShearEnPhase}) with the full $k$-dependent phase
stiffness (\ref{PhStiff-k}). Within these approximations the
energy (\ref{splitLDen}) is replaced by
\begin{equation}
F[\phi_{r},\phi_{v},\mathbf{A}]=F_{r}[\phi_{r},\mathbf{A}_{\perp}
]+F_{v-t}[\phi_{v}]+\tilde{F}_{J}[\phi_{r}+\phi_{v},A_{z}]. \label{AnisLonEn}
\end{equation}
Varying the energy with respect to $\mathbf{A}$, we obtain
\begin{subequations}
\begin{align}
\frac{\Phi_{0}}{2\pi}\mathbf{\nabla}_{\perp}\phi_{r}-\mathbf{A}_{\perp
}+\lambda^{2}\mathbf{\nabla}^{2}\mathbf{A}_{\perp}  &  =0, \label{VectPotPerp}
\\
\frac{\Phi_{0}}{2\pi}\mathbf{\nabla}_{z}\phi-A_{z}+\lambda_{c}^{2}
\mathbf{\nabla}^{2}A_{z}  &  =0. \label{VectPotz}
\end{align}
\end{subequations}
It is convenient to perform the analysis of the large-scale
behavior in $\mathbf{k}$-space. Solving linear equations
(\ref{VectPotPerp}) and (\ref{VectPotz}) using Fourier transform,
\begin{subequations}
\label{VecPot}
\begin{align}
\mathbf{A}_{\perp}  &  =\frac{\Phi_{0}}{2\pi}\frac{\mathbf{\nabla}_{\perp}
\phi_{r}}{1+\lambda^{2}k^{2}},\\
A_{z}  &  =\frac{\Phi_{0}}{2\pi}\frac{\mathbf{\nabla}_{z}\phi}{1+\lambda
_{c}^{2}k^{2}},
\end{align}
\end{subequations}
and excluding $\mathbf{A}$, we express the energy in terms of phases
\begin{widetext}
\[
F=\int\frac{d^{3}\mathbf{k}}{(2\pi)^{3}}\left[  \frac{J}{2s}\frac{\lambda
^{2}k^{2}}{1+\lambda^{2}k^{2}}\left(  \mathbf{\nabla}_{\perp}\phi_{r}\right)
^{2}+\frac{J_{v}(\mathbf{k})}{2s}\left(  \mathbf{\nabla}_{\perp}\phi
_{v}\right)  ^{2}+\frac{sE_{J}}{2}\frac{\lambda_{c}^{2}k^{2}}{1+\lambda
_{c}^{2}k^{2}}\left(  \mathbf{\nabla}_{z}\phi\right)  ^{2}\right]  .
\]
Following the procedure of the previous Section, we eliminate
$\phi_{r}$, minimize the energy with respect to $\phi_{v}$, and
obtain the energy in terms of the total phase
\begin{equation}
F=\int\frac{d^{3}\mathbf{k}}{(2\pi)^{3}}\left[
\frac{J_{\mathrm{eff} }(\mathbf{k)}}{2s}\left(
\mathbf{\nabla}_{\perp}\phi\right)  ^{2}
+\frac{sE_{J}(\mathbf{k})}{2}\left(
\mathbf{\nabla}_{z}\phi\right)
^{2}\right]  . \label{AnLonPhEn}
\end{equation}
where the effective phase stiffness, $J_{\mathrm{eff}}(\mathbf{k)}$, and the
effective Josephson energy are given by
\begin{subequations}
\begin{align}
J_{\mathrm{eff}}^{-1}(\mathbf{k})  &  =J^{-1}\frac{1+\lambda^{2}k^{2}}
{\lambda^{2}k^{2}}+J_{v}^{-1}(\mathbf{k})\\
E_{J}(\mathbf{k})  &  =E_{J}\frac{\lambda_{c}^{2}k^{2}}{1+\lambda_{c}^{2}
k^{2}}
\end{align}
In the case of the JV, minimization with respect to the phases has
to be done with the topological constrain,
$\mathbf{\nabla}_{z}\nabla_{y}\phi-\mathbf{\nabla
}_{y}\nabla_{z}\phi=2\pi\delta(y)\delta(z)$, which gives
\end{subequations}
\begin{subequations}
\label{phase_grad}
\begin{align}
\mathbf{\nabla}_{z}\phi &  =\frac{2\pi
ik_{y}J_{\mathrm{eff}}(\mathbf{k}
)}{J_{\mathrm{eff}}(\mathbf{k})k_{y}^{2}+s^{2}E_{J}(\mathbf{k})k_{z}^{2}}=\frac{2\pi
ik_{y}\left(  1+\lambda_{c}^{2}k^{2}\right)
}{k^{2}\left(  1+\lambda_{c}^{2}k_{y}^{2}+\lambda^{2}k_{z}^{2}(1+w(\mathbf{k}
)\right)  },\\
\mathbf{\nabla}_{y}\phi &  =-\frac{2\pi
ik_{z}s^{2}E_{J}(\mathbf{k}
)}{J_{\mathrm{eff}}(\mathbf{k})k_{y}^{2}+s^{2}E_{J}(\mathbf{k})k_{z}^{2}}=
-\frac{2\pi ik_{z}\left(  1+\lambda^{2}k^{2}\right) }{k^{2}\left(
1+\lambda_{c}^{2}k_{y}^{2}+\lambda^{2}k_{z}^{2}(1+w(\mathbf{k}
))\right)  },
\end{align}
\end{subequations}
and
\begin{align}
\mathcal{E}_{JV}  &  =\frac{1}{2}\int d^{2}\mathbf{k}\frac{sE_{J}
(\mathbf{k})J_{\mathrm{eff}}(\mathbf{k})}{J_{\mathrm{eff}}(\mathbf{k}
)k_{y}^{2}+s^{2}E_{J}(\mathbf{k})k_{z}^{2}}\nonumber\\
&  =\frac{J}{2s}\int \frac{d^{2}\mathbf{k}}{
\lambda^{-2}+\gamma^{2}k_{y} ^{2}+k_{z}^{2}(1+w(\mathbf{k}))
 }  \label{JVEnergyGeneral}
\end{align}
with
\begin{equation}
w(\mathbf{k})=J/J_{v}(\mathbf{k})=\frac{2h}{\lambda^{2}k_{y}^{2}/2+\ln\left(
1+k_{z}^{2}r_{cut}^{2}\right)  } \label{w_k}
\end{equation}
and $h\equiv 4\pi n_{v}\lambda^{2}$. The integration has to be cut
at $k_{z}\sim \pi/s$. In addition, integration with respect to
$k_{y}$ is typically determined by $k_{y}\sim k_{z}/\gamma$ so
that one can neglect in $w(\mathbf{k})$ the term
$\lambda^{2}k_{y}^{2}/2\sim\lambda^{2}k_{z}^{2}/2\gamma^{2}$,
coming from the shear energy, in comparison with the tilt energy
term $\ln\left(  1+k_{z} ^{2}r_{cut}^{2}\right)$ and the JV energy
reduces to
\begin{align}
\mathcal{E}_{JV}  &  \approx\frac{J}{2s}\int d^{2}\mathbf{k}\left(
\lambda^{-2}+\gamma^{2}k_{y}^{2}+k_{z}^{2}+\frac{2hk_{z}^{2}}{\ln\left(
1+k_{z}^{2}r_{cut}^{2}\right)  }\right)  ^{-1}\nonumber\\
&  =\frac{\pi J}{\gamma s}\int_{0}^{k_{c}}\frac{dk_{z}}{\sqrt{\lambda
^{-2}+k_{z}^{2}\left(  1+2h/\ln\left(  1+r_{cut}^{2}/\left(  k_{z}^{-2}
+r_{w}^{2}\right)  \right)  \right)  }}. \label{EnJVLargeSc}
\end{align}
\end{widetext}
A similar formula has been derived in Ref.\ \onlinecite{Savel01}.\
This formula shows that the small-$k_{z}$ logarithmic divergence
in the integral cuts off at $k_{z}=\max(\lambda^{-1},a^{-1})$. To
reproduce JV\ energy at $B_{z}=0$ the upper cutoff has to be
chosen as $k_{c}=2.36/s$.

Lets consider in more detail the case of a large $c$ axis field
$h\gg1$, where JV structure is strongly renormalized by the dense
pancake lattice. Formula for the JV energy simplifies in this
limit to
\[
\mathcal{E}_{JV}\approx\frac{\pi J}{\gamma
s\sqrt{2h}}\int_{0}^{k_{c}} \sqrt{\ln\left(  1+0.05a^{2}/\left(
k_{z}^{-2}+r_{w}^{2}\right)  \right) }\frac{dk_{z}}{k_{z}}.
\]
To estimate this integral, we split the integration region into two intervals,
$1/r_{w}\lesssim k_{z}\lesssim\pi/s$ and $\pi/a\lesssim k_{z}\lesssim1/r_{w}$,
and obtain
\[
\mathcal{E}_{JV}\!\approx\!\frac{\pi J}{\gamma s\sqrt{h}}\left(\!
\sqrt{\ln\!\left( \frac{0.2a}{r_{w}}\right)
}\!\ln\!\frac{2.4r_{w}}{s}\!+\!\frac{2}{3}\left[ \ln\!\left(
\frac{0.2a}{r_{w}}\right) \! \right]  ^{3/2}\!\right).
\]
Note that the long-range contribution to the energy scales as a
logarithm to the power $3/2$.

\begin{widetext}
\subsubsection{\label{Sec:MagField} Magnetic field of Josephson vortex}

Using Eqs.\ (\ref{phase_grad}) and (\ref{VecPot}) we obtain for
the JV magnetic field (see also Ref.\ \onlinecite{Savel01})
\begin{equation}
B_{x}(\mathbf{k})=\frac{\Phi_{0}}{1+\lambda_{c}^{2}k_{y}^{2}+\lambda^{2}
k_{z}^{2}(1+w(\mathbf{k}))},\label{Bx_k}
\end{equation}
where $w(\mathbf{k})$ is given by Eq.\ (\ref{w_k}). Let us
consider the case of large magnetic fields $B>B_{\lambda}$. In a
wide region, $\lambda<\sqrt{y^{2}+(\gamma z)^{2}}<\lambda_{c}$,
the magnetic field in real space is approximately given by
\begin{align}
B_{x}(y,z)  &
\approx\Phi_{0}\int\frac{dk_{y}dk_{z}}{4\pi^{2}}\exp\left(
ik_{y}y+ik_{z}z\right)  \left(
\lambda_{c}^{2}k_{y}^{2}+\frac{2\lambda
^{2}hk_{z}^{2}}{\ln\left(  1+a^{2}k_{z}^{2}/20\right)  }\right)  ^{-1} \label{Bxy}\\
&
=\frac{\Phi_{0}}{2\pi\lambda\lambda_{c}\sqrt{2h}}\int_{0}^{\infty}
dk_{z}\frac{\sqrt{\ln\left(  1+a^{2}k_{z}^{2}/20\right)
}}{k_{z}}\exp\left(
-\frac{\sqrt{2h}k_{z}|y|}{\gamma\sqrt{\ln\left(
1+a^{2}k_{z}^{2}/20\right) }}\right)  \cos\left(
k_{z}z\right)\nonumber
\end{align}
\end{widetext}
One can estimate from this expression the JV maximum field as
\begin{equation}
B_{x}(0,0)\approx\frac{\Phi_{0}}{3\pi\lambda\lambda_{c}\sqrt{h}}\left(
\ln\frac{a}{z_{c}}\right)  ^{3/2} \label{Bmax}
\end{equation}
and this field decays at the scale $\sim a/4.5$ in the $z$
direction and at the scale $\gamma a^{2}/20\lambda$ in the $y$
direction. The magnetic flux concentrated at this region is
estimated as $\Phi\approx\Phi_{0}/(1+2.8h^{2} )$. The residual
flux, $\Phi_{0}-\Phi$, is distributed over the pancake lattice at
much larger distances.

Due to the elasticity of the pancake lattice, the behavior at
large distances is very unusual. The limiting expression for
$B_{x}(\mathbf{k})$ at $\mathbf{k}\rightarrow0$ is given by
\begin{equation}
B_{x}(\mathbf{k})\approx\Phi_{0}\left(  1+\frac{hk_{z}^{2}
}{k_{y}^{2}/4+k_{z}^{2}/(2.8h)}\right)  ^{-1}\label{Bx_small_k}
\end{equation}
Formally, the total flux of JV is given by the limit
\[
\Phi=\lim_{\mathbf{k}\rightarrow0}B_{x}(\mathbf{k})
\]
However this limit depends on the order of limits
$\lim_{k_{y}\rightarrow0}$
and $\lim_{k_{z}\rightarrow0}$
\begin{align*}
\lim_{k_{z}\rightarrow0}\lim_{k_{y}\rightarrow0}B_{x}(\mathbf{k})
&=\frac{\Phi_{0}}{1+2.8h^{2}}\approx \frac{\Phi_{0}}{2.8h^{2}}\\
\lim_{k_{y}\rightarrow0}\lim_{k_{z}\rightarrow0}B_{x}(\mathbf{k})  & =\Phi_{0}
\end{align*}
This apparent paradox can be resolved by calculating the field
distribution in the real space
\begin{align}
B_{x}(y,z)&\approx\frac{1}{2}\left(  1+\frac{1}{2.8h^{2}}\right)
\Phi
_{0}\delta(\mathbf{r})\\
&-\left(  1-\frac{1}{2.8h^{2}}\right)  \frac{\Phi_{0}
}{\pi\sqrt{h}}\frac{4y^{2}-z^{2}/h}{\left(  4y^{2}+z^{2}/h\right)
^{2}},\nonumber \label{B_larger}
\end{align}
This expression clearly shows that the screening is incomplete:
the field at large scales has a slowly decaying $1/r^{2}$ tail.
The magnetic flux through the
large size box $L_{y}\times L_{z}$ is given by
\[
\Phi(L_{y},L_{z})=\Phi_{0}\left(  1-\left(
1-\frac{1}{2.8h^{2}}\right) \frac{2}{\pi}\arctan\left(
\frac{2\sqrt{h}L_{y}}{L_{z}}\right)  \right)
\]
and the limiting value of the total flux at
$L_{y},L_{z}\rightarrow \infty$ depends on the aspect ratio
$L_{y}/L_{z}$.

\begin{figure}
[ptb]
\begin{center}
\includegraphics[clip,width=3.2in ] {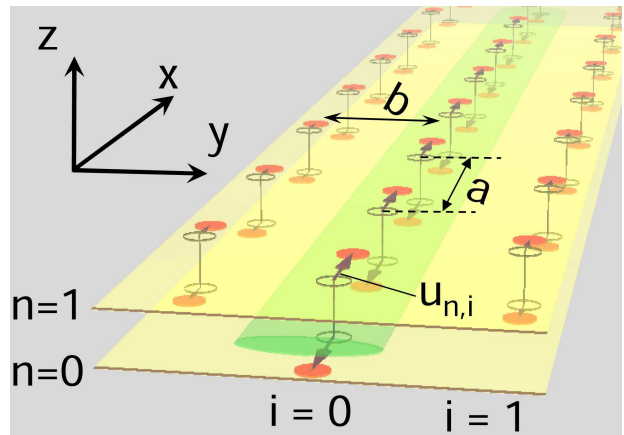}
\caption{Displacements of the pancake rows in the JV core.}
\label{Fig-JVrow_displ}
\end{center}
\end{figure}

\subsection{\label{Sec:QuantCore}Quantitative analysis of the core structure}

The simple ``effective phase stiffness'' approximation, described
in Section \ref{Sec:EffPhStiff}, is only valid if $\gamma$ is
significantly larger than $\lambda/s$. In BSCCO, even at low
temperatures, $\gamma$ is at most $3-4$ times larger than
$\lambda/s$. Moreover, it always approaches $\lambda/s $ at
$T\rightarrow T_{c}$. In this section we extend our analysis to
the region $\gamma\sim\lambda/s$. We consider JV structure at low
temperatures and not very small $c$ axis field,
$B_{z}>\Phi_{0}/(\gamma s)^{2}$. The structure of JV core is
completely determined by the displacements of pancake vortices and
phase distribution. The equilibrium pancake displacements depend
only on the layer index and on coordinate, perpendicular to the
direction of the vortex (see Fig.\ \ref{Fig-JVrow_displ}).
Therefore, the energy can be expressed in terms of the
displacements of the vortex rows $u_{n,i}$. Different
representations for the magnetic interaction between the vortex
rows $U_{Mr}(u_{n,i}-u_{m,j},n-m)$ are considered in Appendix
\ref{App-MagIntRows}. We will operate with the phase perturbation
$\phi_{n}(\mathbf{r})$ with respect to equilibrium phase
distribution of the perfectly aligned pancake crystal. We split
this phase into the contribution, averaged over the JV direction
($x$ axis), $\bar{\phi}_{n}(y)$, and the oscillating in the $x$
direction contribution, $\tilde{\phi}_{n}(x,y)$. Pancake
displacements induce jumps of the average phase at the coordinates
of the vortex rows $Y_{i}$, $\bar{\phi
}_{n}(Y_{i}+0)-\bar{\phi}_{n}(Y_{i}-0)=2\pi u_{n,i}/a$. The
oscillating phase induced by the row displacements becomes
negligible already at the neighboring row. This allows us to
separate the local contribution to the Josephson energy coming
from $\tilde{\phi}_{n}(x,y)$ (see Appendix \ref{AppEJosc}) and
reduce initially three-dimensional problem to the two-dimensional
problem of finding the average phase and row displacements.
Further on we operate only with the averaged phase and skip the
accent ``$-$'' in the notation $\bar{\phi}_{n}(y)$. We again split
the total phase into the continuous regular phase, $\phi_{rn}(y)$,
and the vortex phase, $\phi_{vn}(y)$, $\phi
_{n}(y)=\phi_{rn}(y)+\phi_{vn}(y;u_{n,i})$. The vortex phase is
composed of jumps at the row positions $Y_i$,
\begin{equation}
\phi_{vn}(y;u_{n,i})=-\frac{2\pi}{a}\sum_{i}u_{n,i}\Theta\left(
Y_{i}-y\right)  , \label{VorPhase-discr}
\end{equation}
where $\Theta\left(  y\right)  $ is the step-function
($\Theta\left( y\right) =1$ ($0$) at $y>0$ ($y<0$)). In the
effective phase stiffness approach of Section \ref{Sec:EffPhStiff}
we used a coarse-grained continuous approximation for this phase.
We neglect $x$-dependent contribution in the regular phase, which
is small at $B>\Phi_{0}/(\gamma s)^{2}$. Collecting relevant
energy contributions, we now write the energy per unit length in
terms of the regular phase, $\phi_{rn}(y)$, and the row
displacements, $u_{n,i}$,
\begin{align}
\mathcal{E}_{J}  &  \!=\!\sum_{n}\!\int \! dy\!\left[\!
\frac{J}{2}\left(\! \frac {d\phi_{rn}}{dy}\!\right)
^{2}+E_{J}\left( 1-\cos\left( \phi_{n+1}-\phi
_{n}\right)  \right)  \!\right] \nonumber\\
&  +\frac{1}{2}\sum_{n,m,i,j}\tilde{U}_{Mr}(u_{n,i}-u_{m,j},Y_{i,j}
,n-m)\nonumber\\
&
+\sum_{n,i}\mathcal{E}_{Josc}(u_{n+1,i}-u_{n,i},\phi_{n+1,i}-\phi_{n,i})
\label{EnRowsPhase}
\end{align}
where
\begin{itemize}
\item $\phi_{n}(y)\equiv\phi_{rn}(y)+\phi_{vn}(y;u_{n,i})$ is the
total phase;
\item $U_{Mr}(x_{n,i}-x_{m,j},Y_{i,j},n-m)$ is the
magnetic interaction between the vortex rows separated by distance
$Y_{i,j}=Y_{i}-Y_{j}=b(i-j)$ (see Appendix \ref{App-MagIntRows}),
\begin{align*}
\tilde{U}_{Mr}(u_{n,i}-u_{m,j},\ldots) \equiv &U_{Mr}(X_{i-j}
+u_{n,i} -u_{m,j},\ldots)\\
&-U_{Mr}( X_{i-j},\ldots).
\end{align*}
is the variation of this interaction caused by pancake row
displacements,  $X_{i}=0$ for even $i$ and $X_{i}=a/2$ for odd
$i$.
\newline $\tilde{U}_{Mr}(x,y,n)$ is periodic with
respect to $x$,\newline
$\tilde{U}_{Mr}(x+a,y,n)=\tilde{U}_{Mr}(x,y,n)$; \item
$\mathcal{E}_{Josc}(u_{n+1,i}-u_{n,i},\phi_{n+1,i}-\phi_{n,i})$ is
the local Josephson energy due to the oscillating component of the
phase difference (see Appendix \ref{AppEJosc}) with
\[
\phi_{n,i}\equiv\phi_{rn}(Y_{i})+\pi\frac{u_{n,i}}{a}-\frac{2\pi}{a}\sum
_{j>i}u_{n,j}\Theta\left(  Y_{j}-y\right)
\]
being the external phase at $i$-th rows and $n$-th layer.
\end{itemize}
The energy (\ref{EnRowsPhase}) describes JV\ structure at
distances $r_{\perp}\ll \lambda_{c}$ and $z\ll a,\lambda$ from its
center.

To facilitate calculations we introduce the reduced coordinates
\[
\tilde{y}=\frac{y}{\gamma s},\; v_{ni}=\frac{u_{ni}}{a},
\]
and represent the energies in the scaling form. Magnetic interaction between
the rows we represent as
\[
U_{Mr}(x,y,n)=\frac{\pi Ja}{\lambda^{2}}\mathcal{V}_{Mr}(\frac{x}{a},\frac
{y}{a},n),
\]
\begin{widetext}
where
\begin{align*}
\mathcal{V}_{Mr}(x,y,n)  &  =-\frac{\lambda^{2}}{a^{2}}\left(  \delta
_{n}-\frac{s}{2\lambda}\exp\left(  -\frac{s|n|}{\lambda}\right)  \right)
\ln\left[  1-2\cos2\pi x\exp\left(  -2\pi|y|\right)  +\exp\left(
-4\pi|y|\right)  \right] \\
&  +\frac{s\lambda}{2a^{2}}\sum_{m=-\infty}^\infty u\left(
\frac{\sqrt{y^{2}+(x-m)^{2}}
}{\lambda/a},\frac{s|n|}{\lambda}\right)
\end{align*}
and $u(r,z)$ is defined by Eq.\ (\ref{urz}). Note that at
$x,y\rightarrow 0$ $\mathcal{V}_{Mr}(x,y,n)$ remains finite for
$n\neq 0$ because, logarithmic divergency in the first term is
compensated by logarithmic divergency of the $m=0$ term in the
sum. At $\rho^{2}\equiv x^{2} +y^{2}\rightarrow0$, using
asymptotics $u(\rho,z)\approx\exp(-z)\left( -\gamma_{E}-\ln\left(
\frac{\rho^{2}}{2z}\right)  \right)  +\exp (z)E_{1}\left(
2z\right)  $ with $\gamma_{E}=0.5772$, we obtain the limiting
value of $\mathcal{V}_{Mr}(x,y,n)$ at $n\neq0$
\[
\mathcal{V}_{Mr}(0,0,n)  =\frac{s\lambda}{2a^{2}}\left [\exp\left(
-\frac{s|n|}{\lambda}\right)  \left(  \ln\left[
\frac{8\pi^{2}s|n|\lambda }{a^{2}}\right]  -\gamma_{E}\right)
  +\exp\left(  \frac{s|n|}{\lambda}\right)
E_{1}\left( \frac{2s|n|} {\lambda}\right)
+2\sum_{m=1}^{\infty}u\left
(\frac{am}{\lambda},\frac{s|n|}{\lambda}\right)\right ].
\]
The
local Josephson energy can be represented as
\[
\mathcal{E}_{Josc}(u,\phi)=2E_{J}a\cos(\phi)\mathcal{J}(u/a),
\]
where $\mathcal{J}(v)$ is dimensionless function,
$\mathcal{J}(v)\approx \frac{\pi}{4}v^{2}\ln\frac{0.39}{v}$ at
$v\ll1$ (see Appendix \ref{AppEJosc}). In reduced units the total
energy takes the form
\begin{align}
\mathcal{E}_{J}/\varepsilon_{J0}  &  =\sum_{n}\int d\tilde{y}\left[  \frac
{1}{2}\left(  \frac{d\phi_{rn}}{d\tilde{y}}\right)  ^{2}+1-\cos\left(
\phi_{n+1}-\phi_{n}\right)  \right] \nonumber\\
&  +\frac{\pi\gamma sa}{2\lambda^{2}}\sum_{n,m,i,j}\mathcal{\tilde{V}}
_{Mr}(v_{n,i}-v_{m,j},\frac{\tilde{Y}_{i,j}}{a_{\gamma}},n-m)\nonumber\\
&  +\frac{2a}{\gamma s}\sum_{n,i}\mathcal{J}(v_{n+1,i}-v_{n,i})\cos
(\phi_{n+1,i}-\phi_{n,i}) \label{RedEn}
\end{align}
where $\varepsilon_{J0}\equiv E_{J}\gamma s\equiv J/\gamma s$ is
the JV energy scale, $\phi_{n}(y)=\phi_{rn}(y)-2\pi\sum_{i}
v_{n,i}\Theta\left( Y_{i}-y\right)  $ and $a_{\gamma}\equiv
a/\gamma s$. Varying the energy, we obtain equations for $v_{n,i}$
and $\phi_{n}$
\begin{subequations}
\begin{align}
&  \nabla_{y}\phi_{rn}(Y_{i})+\frac{\gamma sa}{2\lambda^{2}}\sum
_{m,j}\mathcal{F}_{Mr}(v_{n,i}-v_{m,j},\frac{\tilde{Y}_{i,j}}{a_{\gamma}
},n-m)+\frac{a}{\pi\gamma s}\sum_{\delta=\pm1}\cos(\phi_{n,i}-\phi
_{n+\delta,i})\mathcal{F}_{J}(v_{n,i}-v_{n+\delta,i})=0,\label{RedEqU}\\
&  \frac{d^{2}\phi_{rn}}{d\tilde{y}^{2}}+\sin\left(  \phi_{n+1}-\phi
_{n}\right)  -\sin\left(  \phi_{n}-\phi_{n-1}\right)  =0. \label{RedEqPhase}
\end{align}
\end{subequations}
Here
$\mathcal{F}_{Mr}(x,y,n)\equiv-\nabla_{x}\mathcal{V}_{Mr}(x,y,n)$
is the magnetic interaction force between the vortex rows
\begin{align*}
\mathcal{F}_{Mr}(x,y,n)  &  =\frac{2\pi\lambda^{2}}{a^{2}}\left(
\delta _{n}-\frac{s}{2\lambda}\exp\left(
-\frac{s|n|}{\lambda}\right)  \right)
\frac{\sin2\pi x}{\cosh2\pi|y|-\cos2\pi x}\\
&  +\frac{s\lambda}{a^{2}}\sum_{m}\frac{x-m}{\left(  x-m\right)
^{2}+y^{2} }\exp\left(  -\frac{\sqrt{\left(  x-m\right)
^{2}+y^{2}+z^{2}}}{\lambda /a}\right)  .
\end{align*}
and $\mathcal{F}_{J}(v,\phi)=-\partial\mathcal{J}(v)/\partial v\approx
-(\pi/2)v\ln(0.235/v)$ at $v\ll1$. The derivative of the regular phase has
jumps at the positions of the rows
\begin{equation}
\frac{d\phi_{rn}}{d\tilde{y}}(\tilde{Y}_{i}+0)-\frac{d\phi_{rn}}{d\tilde{y}
}(\tilde{Y}_{i}-0)=\frac{2a}{\gamma s}\sum_{\delta=\pm1}\mathcal{J}
(v_{n+\delta,i}-v_{n,i})\sin(\phi_{n+\delta,i}-\phi_{n,i}). \label{DerJump}
\end{equation}
\end{widetext}
To find JV structure at low temperatures one has to solve Eqs. (\ref{RedEqU}),
(\ref{RedEqPhase}), and (\ref{DerJump}) with condition (\ref{JVcond}).

Let us consider in more detail magnetic interactions between vortex rows,
i.e., the term with $\mathcal{F}_{Mr}$ in Eq.\ (\ref{RedEqU}). Firstly, one
can observe that the dominating contributions to the sum over the layer index
$m$ and row index $j$ come from rows in the same layer, $m=n$, and rows in the
same of stacks, $j=i$. The former sum determines the shear stiffness, while
the latter one determines the magnetic tilt stiffness.
\begin{equation}
\sum_{m,j}\mathcal{F}_{Mr}(v_{n,i}-v_{m,j},\frac{\tilde{Y}_{i,j}}{a_{\gamma}
},n-m)\approx f_{\mathrm{shear}}\left[  v_{n,i}\right]  +f_{\mathrm{tilt}
}\left[  v_{n,i}\right]  \label{FMr-split}
\end{equation}
with
\begin{align}
f_{\mathrm{shear}}\left[  v_{n,i}\right]   &  =\sum_{j\neq i}\mathcal{F}
_{Mr}(v_{n,i}-v_{n,j},\frac{\tilde{Y}_{i,j}}{a_{\gamma}},0)\label{fshear}\\
f_{\mathrm{tilt}}\left[  v_{n,i}\right]   &  =\sum_{m\neq n}\mathcal{F}
_{Mr}(v_{n,i}-v_{m,i},0,n-m) \label{ftilt}
\end{align}
The sum over the rows in $f_{\mathrm{shear}}\left[  v_{n,i}\right]
$ converges very fast and effectively is determined by the first
two neighboring rows. Note that skipping the terms with $m\neq n$
in $f_{\mathrm{shear} }\left[  v_{n,i}\right]  $ is completely
justified in the limit $a<\lambda$ but leads to overestimation of
the shear energy in the limit $a>\lambda$. However in this limit
the shear energy has already a very weak influence on JV
properties. The sum over the layers in $f_{\mathrm{tilt}}\left[
v_{n,i}\right]  $ is determined by large number of the layers of
the order of $a/s$ or $\lambda/s$. If we consider layer $n$ close
to the JV\ core, then the interaction force with row in the layer
$m$, in the same stack with $n\ll m\ll a/s,\lambda/s$ is given by
$\mathcal{F}_{Mr}\left(  v_{n,i}-v_{m,i} ,0,n-m\right)
\approx-(v_{n,i}-v_{m,i})/2(m-n)$. Interactions with remote layers
give large contributions even if displacements in these layers are
small, $v_{m,i}\ll v_{n,i}$. A useful trick to treat this
situation is to separate interaction of a given pancake row with
the \emph{aligned} stack  pancake rows:
\[
f_{\mathrm{tilt}}\left[  v_{n,i}\right]  =\sum_{m\neq
n}\mathcal{\tilde{F} }_{Mr}\left(  v_{n,i}-v_{m,i},0,n-m\right)
+f_{\mathrm{cage}}\left( v_{n,i}\right)
\]
with
\begin{align*}
&\mathcal{\tilde{F}}_{Mr}\left( v_{n,i}-v_{m,i},0,n-m\right)\\
&=\mathcal{F} _{Mr}\left( v_{n,i}-v_{m,i},0,n-m\right)
-\mathcal{F}_{Mr}\left( v_{n,i},0,n-m\right)
\end{align*}
and
\[
f_{\mathrm{cage}}\left(  v\right)  =\sum_{m\neq0}\mathcal{F}_{Mr}\left(
v,0,m\right)
\]
is the interaction force of a chosen pancake row with the aligned
stack of rows (``cage'' force), for which we derive a useful
representation
\begin{widetext}
\[
f_{\mathrm{cage}}\left(  v\right)  =-\sum_{l=1}^{\infty}\frac{4\pi\sin\left(
2\pi lv\right)  }{\sqrt{(a/\lambda)^{2}+\left(  2\pi l\right)  ^{2}}\left(
\sqrt{(a/\lambda)^{2}+\left(  2\pi l\right)  ^{2}}+2\pi l\right)  }.
\]
At $v\ll1\ll\lambda/a$ this equation gives
$f_{\mathrm{cage}}\left(  v\right) \approx-v\ln(0.433/v)$.
$\mathcal{\tilde{F}}_{Mr}\left(  v_{n,i} -v_{m,i},0,n-m\right)  $
behaves as $v_{m,i}/2(m-n)$ at large $m$ and decays with increase
of $m$ much faster than $\mathcal{F}_{Mr}\left(  v_{n,i}
-v_{m,i},0,n-m\right)  $. The same splitting can be made in the
magnetic coupling energy:
\begin{align*}
\frac{1}{2}\sum_{m\neq n}\mathcal{V}_{Mr}(v_{n}-v_{m},n-m)  &  =\sum
_{|m|>|n|}\left(  \mathcal{V}_{Mr}(v_{n}-v_{m},n-m)-\mathcal{V}_{Mr}
(v_{n},n-m)-\mathcal{V}_{Mr}(v_{m},n-m)\right) \\
&  +\sum_{n}v_{\mathrm{cage}}\left(  v_{n}\right)
\end{align*}
with
\begin{align*}
v_{\mathrm{cage}}\left(  v\right)   &  =\sum_{n\neq0}\mathcal{V}_{Mr}(v,n)\\
&  =\sum_{l=1}^{\infty}\frac{2\left(  1-\cos\left(  2\pi lv\right)  \right)
}{l\sqrt{(a/\lambda)^{2}+\left(  2\pi l\right)  ^{2}}\left(  \sqrt
{(a/\lambda)^{2}+\left(  2\pi l\right)  ^{2}}+2\pi l\right)  }
\end{align*}
\end{widetext}
This function has simple asymptotics at $v\ll1\ll\lambda/a$,
$v_{\mathrm{cage}}\left( v\right) \approx (v^{2}/2)\ln(0.713/v)$.

We demonstrate now that in the limit $\gamma\gg\lambda/s$ Eqs.\
(\ref{RedEqU}) and (\ref{RedEqPhase}) reproduce the JV structure
obtained within the effective phase stiffness approximation. One
can show that in this limit the local Josephson energy influences
weakly the JV structure. We calculate correction to the JV energy
due to this term in the Appendix \ref{App:ContrLocalTerm}. The
dominating contribution to the magnetic interaction between the
pancake rows, $\sum
_{m,j}\mathcal{F}_{Mr}(v_{n,i}-v_{m,j},\frac{Y_{i,j}}{a_{\gamma}},n-m)$,
comes from the tilt force (\ref{ftilt}), which with good accuracy
can be described by the cage force $f_{\mathrm{cage}}\left(
v_{n,i}\right)  $ in the limit of small $v_{n,i}$,
$f_{\mathrm{cage}}\left(  v\right)  \approx-v\ln(C/v)$ with
$C\approx0.433.$ Further estimate shows, that the term
$\sum_{m\neq n}\mathcal{\tilde{F}}_{Mr}\left(
v_{n,i}-v_{m,i},0,n-m\right)  $ in Eq.\ (\ref{ftilt}) amounts to
the replacement of numerical constant $C$ under the logarithm by
slowly changing function of the order of unity. Because of slow
space variations, the discrete row displacements $v_{n,i}$ can be
replaced by the continuous displacement field $v_{n}(y)$. Within
these approximations Eq.\ (\ref{RedEqU}) reduces to
\[
\nabla_{y}\phi_{rn}\approx\frac{\gamma sa}{2\lambda^{2}}v_{n}(y)\ln\frac
{C}{v_{n}(y)}
\]
Replacing $v(y)$ by the vortex phase $\phi_{vn}(y)$ obtained by
coarse-graining of Eq.\ (\ref{VorPhase-discr}), $v_{n}(y)=\left(  b/2\pi\gamma
s\right)  \nabla_{y}\phi_{vn}(y)$, we obtain
\[
\nabla_{y}\phi_{rn}\approx\frac{B_{\lambda}}{B_{z}}\nabla_{y}\phi_{vn}
\]
Note that we replaced $v_{n}(y)$ in $\ln[C/v_{n}(y)]$ by its
typical value and absorbed the logarithmic factor into the
definition of $B_{\lambda} \;$(\ref{VorPhStiff}). From the last
equation we obtain $\phi_{rn} =(B_{\lambda}/B_{z})\phi_{vn}$and
\[
\phi_{n}=\left(  1+\frac{B_{z}}{B_{\lambda}}\right)  \phi_{rn}
\]
Therefore Eq.\ (\ref{RedEqPhase}) reduces to
\[
-\frac{1}{1+\frac{B_{z}}{B_{\lambda}}}\frac{d^{2}\phi_{n}}{dy^{2}}+\sin\left(
\phi_{n}-\phi_{n+1}\right)  +\sin\left(  \phi_{n}-\phi_{n-1}\right)  =0,
\]
which is just dimensionless version of Eq.\ (\ref{PhaseEq}).

\subsubsection{Numerical calculations of JV core structure. Crossover to solitonlike cores.}
\begin{figure}
\begin{center}
\includegraphics[width=2.4in ] {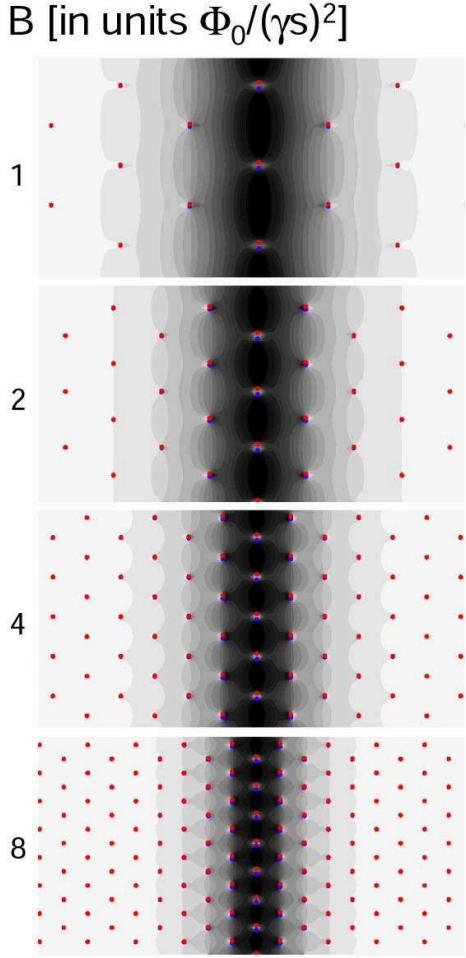}
\caption{Grey level plots of
cosine of the phase difference between two central layers of JV,
$\cos\Theta$, for $\lambda=0.2\gamma s$ and several magnetic
fields (dark regions correspond to $\cos\Theta\sim -1$ and white
regions correspond to $\cos\Theta\sim 1$). The total size of
displayed region in the horizontal direction is $6 \gamma s$. One
can see that in this regime several pancake rows fit inside the
core region. At high fields the size of the core shrinks so that
the number of pancake rows inside the core remains constant.}
\label{Fig-PhaseDifAl02}
\end{center}
\end{figure}
\begin{figure}
[ptb]
\begin{center}
\includegraphics[clip,width=2.5in ] {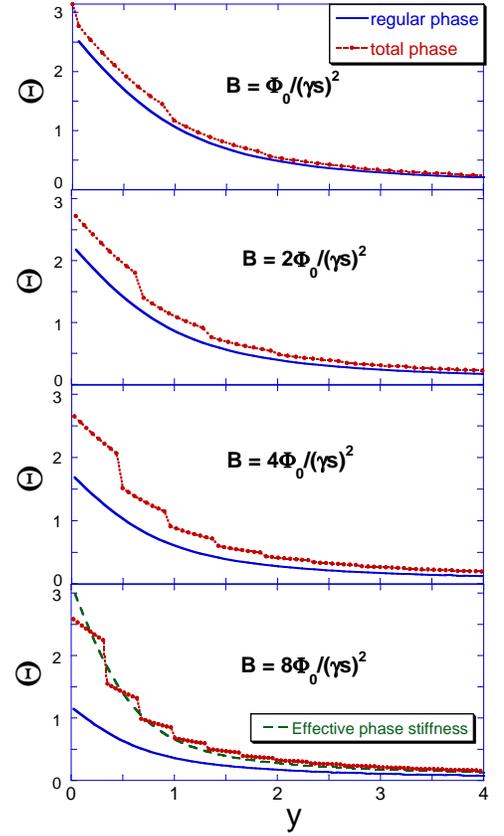}
\caption{Coordinates dependence
of the phase difference between two central layers,
$\Theta=\phi_{1}-\phi_{0}$,  for $\lambda/\gamma s=0.2$ and
different magnetic fields. Circles connected by dotted lines
represent total phase difference, solid lines show contributions
from the regular phase. Jumps of the total phase difference at the
positions of pancake rows are caused by pancake displacement and
represent the vortex phases. In the lower plot dashed line
represents prediction of the ``effective phase stiffness'' model
with
$B_{\lambda}=2.1\Phi_{0}/4\pi\lambda^{2}$.  }
\label{Fig-Phase-yAl02}
\end{center}
\end{figure}
\begin{figure}
[ptbptbptb]
\begin{center}
\includegraphics[clip,width=2.8in ] {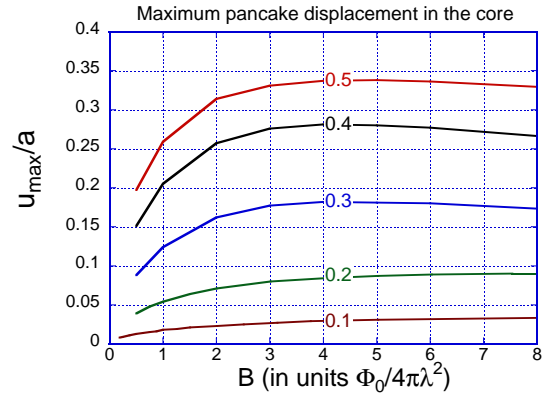}
\caption{Field dependencies of the
maximum pancake displacement in the core at
different ratios $\lambda/\gamma s$ (the curves are labelled by this ratio).}
\label{Fig-vmaxB}
\end{center}
\end{figure}
\bigskip
\begin{figure}
\begin{center}
\includegraphics[clip,width=3.4in ] {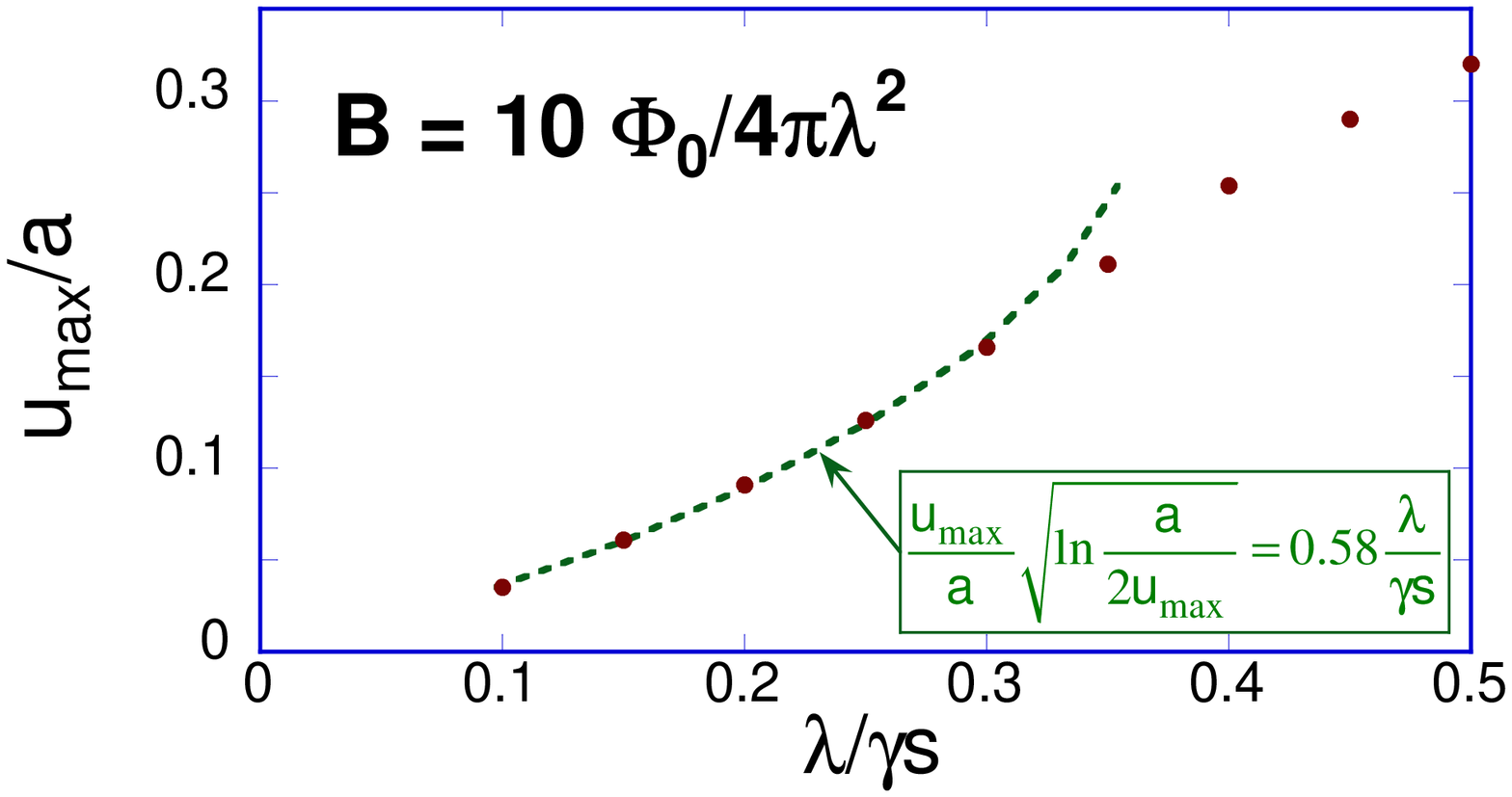}
\caption{Dependence of the maximum
pancake displacement in the core on the ratio $\lambda/\gamma s$
at $B=10\Phi_{0}/4\pi\lambda^{2}$. Dashed line represents
prediction of the ``effective phase stiffness''  model
(\ref{MaxDisplHighB}).} \label{Fig-vmaxPhb10}
\end{center}
\end{figure}

To explore the JV core structure, we solved Eqs.\ (\ref{RedEqU})
and (\ref{RedEqPhase}) numerically for different ratios
$\lambda/\gamma s$ and different magnetic fields. We used a
relaxation technique to find the equilibrium displacements of the
pancake rows and the continuous regular phase. Typically, we
solved equations for $20$ layers and the in-plane region
$0<\tilde{y}<20$.

To test the ``effective phase stiffness'' model and to calculate
uncertain numerical factors, we start from the case of large
anisotropies, $\gamma\gg \lambda/s$. Figure \ref{Fig-PhaseDifAl02}
shows the grey level plots of the cosine of the phase difference
between two central layers of JV, $\cos \Theta $, $\Theta \equiv
\phi_{1}-\phi_{0}\equiv 2\phi_{1}$, for $\lambda =0.2 \gamma s$.
Fig.\ \ref{Fig-Phase-yAl02} shows $y$ dependence of the total
phase difference $\Theta$ and the contribution to this phase
coming from the regular phase for the same parameters. As one can
see, at $B\gtrsim\Phi_{0}/(\gamma s)^{2}$ the core region covers
several pancake rows. At high fields the core size shrinks so that
the number of rows in the core does not change, in agreement with
the ``effective phase stiffness'' model. From Fig.\
\ref{Fig-Phase-yAl02} one can see that the fraction of the regular
phase in the total phase progressively decreases with increase of
magnetic field. For the field $8 \Phi_0/(\gamma s)^2$ we also
plotted $\Theta (y)$ dependence from the ``effective phase
stiffness'' model, assuming $B_\lambda = 2.1
\Phi_0/4\pi\lambda^2$. One can see that the numerically calculated
dependence is reasonably well described by this model.
\begin{figure*}
\begin{center}
\includegraphics[clip,width=5.8in]{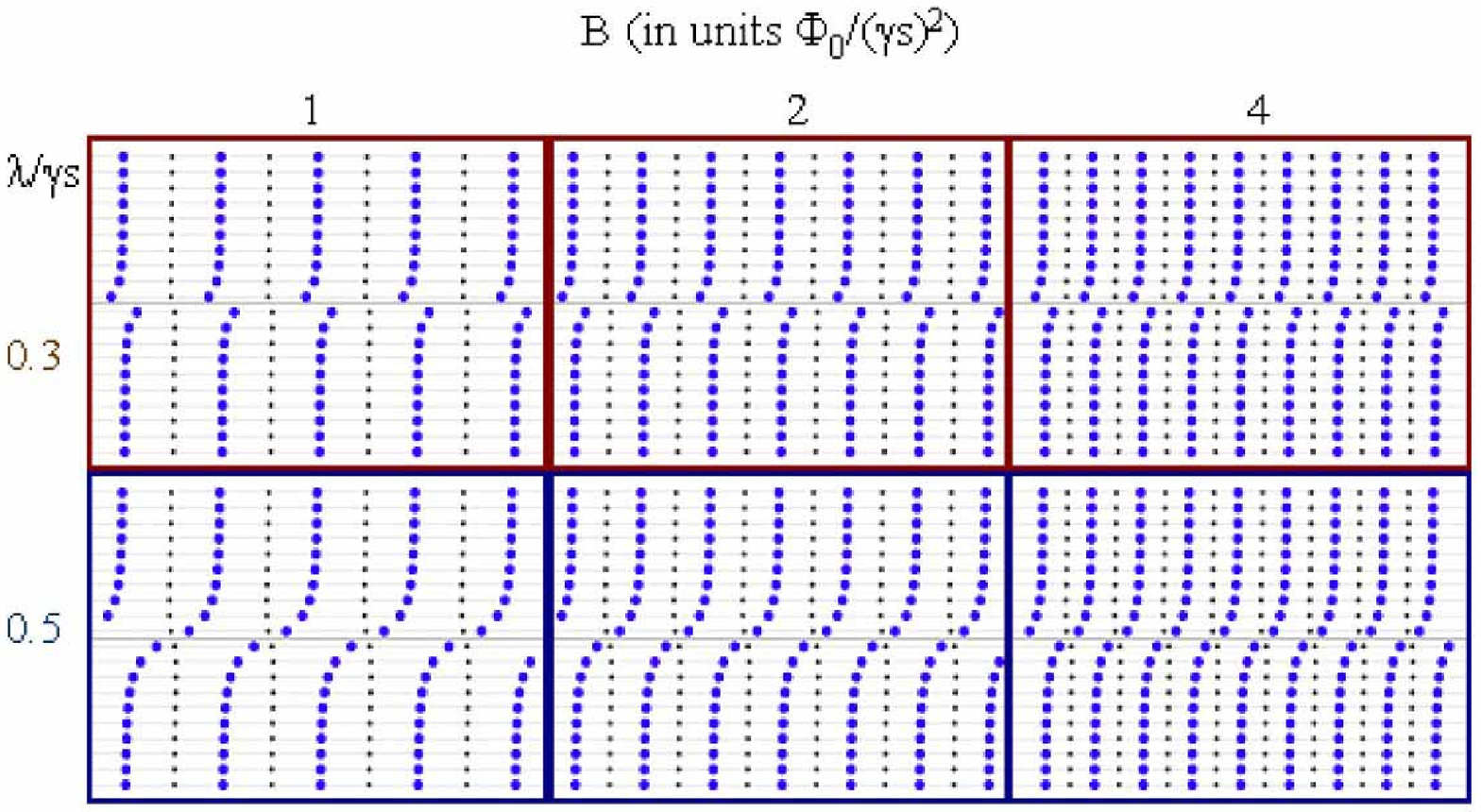}
\caption{Structure of the pancake-stacks row in the center of JV
(big circles) and its neighboring row (small circles)  for
$\lambda/\gamma s=0.3$ and $0.5$ and several values of the
magnetic field.  At $\lambda/\gamma s=0.5$ pancakes in the central
row form lines smoothly transferring between two ideal  lattice
position (solitonlike structure).} \label{Fig-JVdsp}
\end{center}
\end{figure*}
\begin{figure*}
\begin{center}
\includegraphics[width=5.8in]{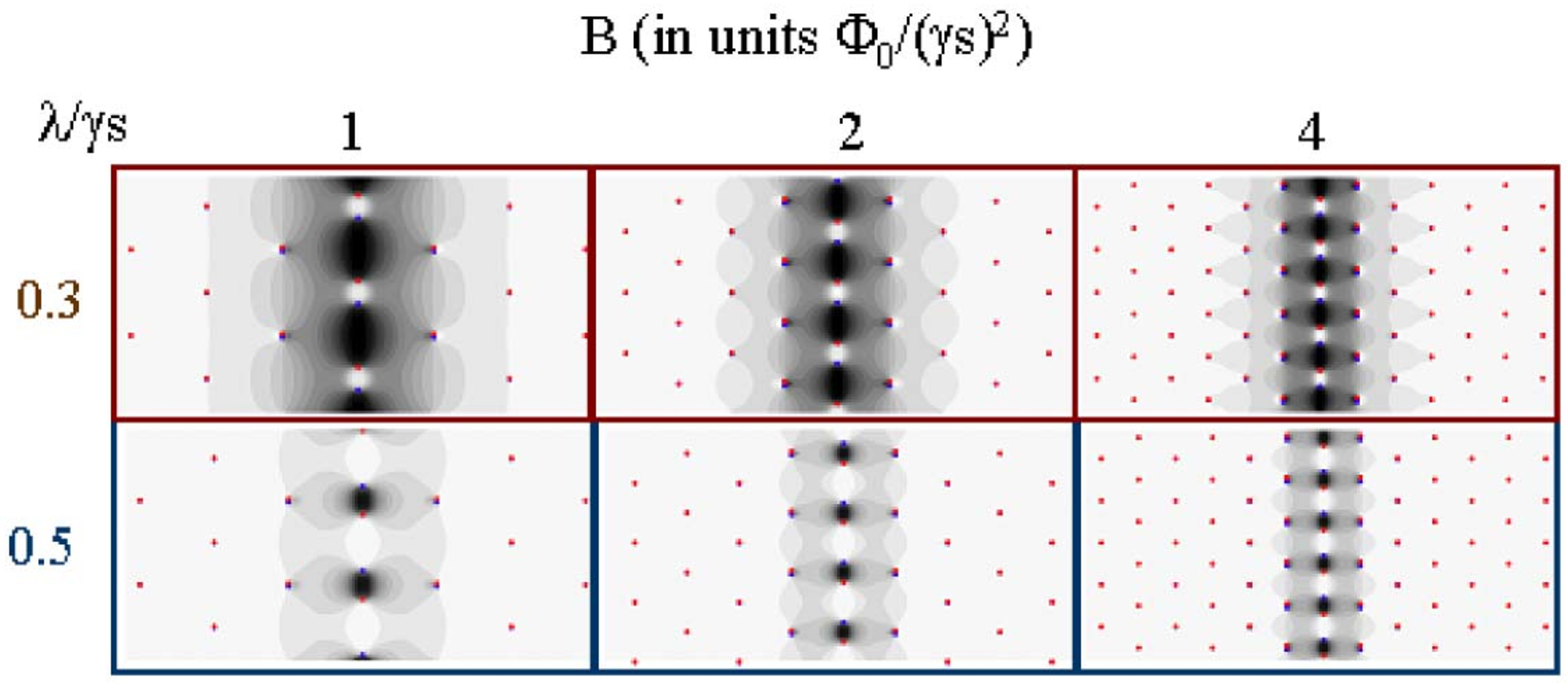}
\caption{Gray-level plots of cosine of interlayer
phase difference between the two central layers of JV for
$\lambda/\gamma s=0.3$ and $0.5$ and different fields. For
$\lambda/\gamma s=0.3$ the JV core covers roughly three pancake
rows, while for $\lambda/\gamma s=0.5$ it shrinks to one pancake
rows. In the second case the regions of suppressed Josephson
energy are practically eliminated.} \label{Fig-JVphase}
\end{center}
\end{figure*}

We now extend study of the core structure to moderate anisotropies
$\sim\lambda/s$. At Fig.\ \ref{Fig-vmaxB} we plot the maximum
pancake displacement $u_{\rm max}$ in the core region normalized
to the lattice constant $a$ as function of magnetic field for
different $\lambda/\gamma s$. The maximum displacement
approximately saturates at a finite fraction of lattice constant
at high field (at high $\lambda/\gamma s$ one can actually observe
a slight decrease of $u_{\rm max}/a$ with field). Fig.\
\ref{Fig-vmaxPhb10} shows dependence of $u_{\rm max}/a$ on
$\lambda/\gamma s$ for fixed field $B=10 \Phi_0 /(\gamma s)^2$.
Dashed line shows prediction of the ``effective phase stiffness''
model given by Eq.\ (\ref{MaxDisplHighB}). One can see that this
equation correctly predicts maximum displacement for
$\lambda/\gamma s<0.35$. Important qualitative change occurs at
$\lambda/\gamma s
> 0.35$, where the maximum displacement $u_{\rm max}(0)$ exceeds
$a/4$. This means that the pancakes initially belonging to the
neighboring stacks become closer than the pancakes belonging to
the same stack. This can be viewed as switching of the vortex
lines in the central layer of JV. This switching is clearly
observed in Fig.\ \ref{Fig-JVdsp}, which shows pancake
displacements in the central row of pancake stacks and its
neighboring row for two values of the ratio $\lambda/\gamma s$,
0.3 and 0.5, and several fields. For $\lambda/\gamma s=0.5$
configuration of the pancake rows in the central stack is very
similar to the classical soliton ("kink") of the stationary
sine-Gordon equation: the stacks smoothly transfer between the two
ideal lattice position in the region of the core. Simplified
approximate description of such solitionlike structure in the case
$\gamma s <\lambda$ is presented below in Sec.\ \ref{Sec:soliton}.
Fig.\ \ref{Fig-JVphase} shows distribution of cosine of interlayer
phase difference between two central layers, $\cos\Theta$. As one
can see, at $\lambda/\gamma s=0.3$ there are still extended
regions of large phase mismatch in the JV core (dark regions),
while for $\lambda/\gamma s=0.5$ these regions are almost
eliminated by large pancake displacements in the core.

\subsection{\label{Sec:soliton}Simple model for solitonlike cores at moderate anisotropies}

In this Section we consider the structure of the JV\ core for
moderate anisotropies $\gamma s<0.5\lambda$ and high fields
$B>\Phi_{0}/4\pi\lambda ^{2}$. Estimate (\ref{MaxDisplHighB}) and
numerical calculations show that at sufficiently small anisotropy,
$\gamma< 0.5\lambda/s$, the maximum displacement in the core
region exceeds a quarter of the lattice spacing. This means that
distance between displaced pancakes belonging to the same vortex
line, $2u_{x,0}(0)$, becomes larger than the distance between
pancakes initially belonging to the neighboring lines,
$a-2u_{x,0}(0)$ . This can be viewed as switching of the vortex
lines in the central layer of JV. At lower anisotropies pancake
stacks in the central row acquire structure, similar to the
soliton of the stationary sine-Gordon model: in the core region
they displace smoothly between the two ideal lattice position (see
Fig.\ \ref{Fig-JVdsp} for $\lambda/\gamma s=0.5$). In such
configuration a strong phase mismatch between the two central
layers is eliminated, which saves the Josephson energy in the core
region. On the other hand, large pancake displacements lead to
greater loss of the magnetic coupling energy. To describe this
soliton structure, we consider a simplified model, in which we (i)
keep only displacements in the central row $v_{n,0}\equiv v_{n}$,
(ii) use the cage approximation for magnetic interactions and
(iii) neglect the shear energy. All these approximations are valid
close to the JV center. At $n>0$ we redefine displacements as
$v_{n}\rightarrow1+v_{n}$. The redefined displacements depend
smoothly on the layer index, $v_{n+1}-v_{n}\ll1$, so that one can
replace the layer index $n$ by continuous variable $z=ns$,
$u_{n}\rightarrow u(z)$, and use elastic approximation for the
Josephson tilt energy:
\begin{equation}
\mathcal{E}_{\mathrm{core}}\!\approx\int\! dz\!\left(  \frac{\pi
E_{J} s\mathcal{L}_{J}}{2a}\left(  \frac{du}{dz}\right)
^{2}+\frac{\pi Ja} {s\lambda^{2}}v_{\mathrm{cage}}\left
(\frac{u}{a}\right )\right),
\end{equation}
where the logarithmic factor $\mathcal{L}_{J}$ is estimated as
$\mathcal{L}_{J}\approx\ln\left[ 0.39(a/s)\left| du/dz\right|
^{-1}\right]  $ and
\[
v_{\mathrm{cage}}(v)\approx\sum_{l=1}^{\infty}\frac{1-\cos\left(  2\pi
lv\right)  }{(2\pi)^{2}l^{3}}
\]
is the magnetic cage. For estimates, we replace $\mathcal{L}_{J}$
by a constant substituting a typical value for $du/dz$ under the
logarithm. The equilibrium reduced displacement $v=u/a$ is
determined by equation
\begin{equation}
-\frac{d^{2}v}{dz^{2}}+\frac{\gamma^{2}}{\mathcal{L}_{J}\lambda^{2}
}v_{\mathrm{cage}}^{\prime}(v)=0. \label{vcoreEq}
\end{equation}
Its solution is implicitly determined by the integral relation
\[
\int_{1/2}^{v}\frac{dv}{\sqrt{v_{\mathrm{cage}}(v)}}=\frac{A\gamma
z} {\lambda}
\]
with $A=\sqrt{\frac{2}{\mathcal{L}_{J}}}$. Therefore a typical size of the
soliton is given by
\[
z_{s}\approx\lambda/\gamma
\]
and the applicability condition of this approach $z_{s}\gg s$ is
equivalent to $\gamma s\ll\lambda$. In fact, an accurate numerical
calculations of the previous section show that the core acquires
the solitonlike structure already at $\gamma s\lesssim2\lambda$.
The core energy is given by
\[
\mathcal{E}_{\mathrm{core}}\approx\frac{a\pi\sqrt{2JE_{J}\mathcal{L}_{J}}
}{\lambda}\int_{0}^{1}dv\sqrt{v_{\mathrm{cage}}(v)}
\]
At $a\ll\lambda$ numerical evaluation of the integral gives
$\int_{0} ^{1}dv\sqrt{v_{\mathrm{cage}}(v)}\approx1.018/2\pi$ and
we obtain
\[
\mathcal{E}_{\mathrm{core}}\approx\sqrt{JE_{J}\frac{\mathcal{L}_{J}}{2}}
\frac{a}{\lambda}.
\]
We also estimate the shear contribution to the periodic potential
$v_{\mathrm{cage}}(v)$
\begin{align*}
v_{\mathrm{shear}}(v)  &  \approx\frac{2\lambda^{2}}{a^{2}}\frac{2\left(
1-\cos2\pi v\right)  \exp\left(  -\sqrt{3}\pi\right)  }{\left(  1+\exp\left(
-\sqrt{3}\pi\right)  \right)  ^{2}}\\
&  \approx0.017\;\frac{\lambda^{2}}{a^{2}}\left(  1-\cos2\pi v\right)  .
\end{align*}
It occurs to be numerically small and only has to be taken into account when
$a$ becomes significantly smaller than $\lambda$.

It is important to note that the described model does not provide
precise soliton structure. At distances $z\gtrsim \lambda/\gamma $
displacements in other pancake rows become comparable with
displacements in the central row. In fact, at distances $z\gg
\lambda/\gamma $ the displacements should cross over to the
regime, described by the ``effective phase stiffness'' model
(\ref{CoreDispl}). Precise description of the soliton structure is
rather complicated and beyond the scope of this paper.

\section{\label{Sec:JVPin}Pinning of Josephson vortex by pancake vortices}

Dynamic properties of JVs can be probed either by applying
transport current along the $c$ direction or by studying
\textit{ac} susceptibility for magnetic field, polarized along the
layers. Although there are numerous experimental indications that
pancake vortices strongly impede motion of JVs,
\cite{LatPhysC91,EnriquezPRB96,HechtfischerPRB97} no quantitative
study (theoretical or experimental) has been done yet. In this
section we consider a pinning force, which is necessary to apply
to the JV to detach it from the pancake vortex crystal. We
consider the simplest case of a dilute pancake lattice,
$a>\lambda$, which allows us to neglect the influence of pancakes
on the JV core, and small concentration of JVs, so that we can
neglect influence of the JV lattice on the pancake crystal (e.g.,
formation of phase-separated states).

\subsection{Pinning by a single pancake-stack row ($a_{0}>\gamma s$)}

\begin{figure}[ptb]
\begin{center}
\includegraphics[width=2.1in ] {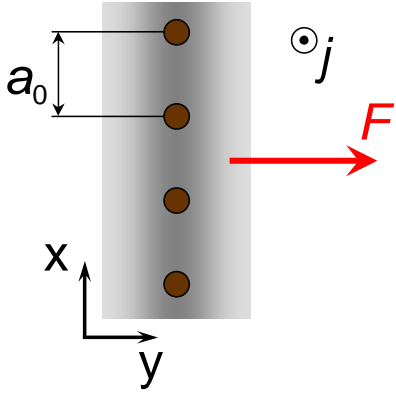}
\caption{Pinning of a Josephson vortex by a single pancake-stack
row.} \label{Fig-SingleRow}
\end{center}
\end{figure}
We consider first the simplest case of an isolated pancake-stack
row crossing JV (see Fig.\ \ref{Fig-SingleRow}) and estimate the
force necessary to detach JV from this stack, assuming that its
position is fixed. The consideration is based on the crossing
energy of JV and isolated pancake stack, calculated in Sec.\
\ref{Sec:CrEn}. Let us calculated first the force necessary to
separate JV from an isolated pancake stack. Using Eq.\
(\ref{CrossEn-y}) we obtain for the force acting on JV from the
pancake stack located at distance $y$ from the center of JV core
\begin{align}
F_{\times}  &  =-\frac{d}{dy}E_{\times}(y)\label{CrossForce}\\
&
\approx\frac{\Phi_{0}^{2}}{4\pi^{2}\gamma^{3}s^{2}\ln(\lambda/u_{1}
(y))}A_{\times}^{\prime}\left(  \frac{y}{\gamma s}\right)
\nonumber
\end{align}
with $A_{\times}^{\prime}\left(  \tilde{y}\right)  \equiv dA_{\times}\left(
\tilde{y}\right)  /d\tilde{y}$. For the maximum force we obtain
\[
F_{\times\max}\approx\frac{C_{f}\Phi_{0}^{2}}{4\pi^{2}\gamma^{3}s^{2}
\ln(\lambda/s\beta_{f})}
\]
with $C_{f}=\max_{\tilde{y}}\left[
A_{\times}^{\prime}(\tilde{y})\right]  $ and $\beta_{f}\sim1$.
Numerical calculation gives $C_{f}\approx1.4$ and the maximum is
located at $y_{f}\approx0.52\gamma s$. The critical current which
detaches JV from the row of pancake stacks with the period $a_{0}$
is given by
\begin{equation}
j_{Jp}=j_{J}\frac{C_{f}\lambda^{2}}{\gamma sa_{0}\ln(\lambda/s\beta_{f})}.
\label{jJp}
\end{equation}
where $j_{J}=c\Phi_{0}/(8\pi^{2}\lambda_{c}^{2}s)$ is the
Josephson current. This expression is valid as long as the lattice
period $a_{0}$ is much larger than the Josephson length $\gamma
s$. Otherwise interaction with several pancake rows has to be
considered, which will be done in the next section.

\subsection{Pinning by dilute pancake lattice ($\lambda<a_{0}<\gamma
s$)}

When several pancake-stack rows fit inside the JV core (but still
$a_{0}>\lambda $) interaction energy of JV per unit length with
the pancake lattice, $E_{Jl}(y)$, can be calculated as a sum of
crossing energies (\ref{CrossEn-y}),
\begin{align}
E_{Jl}(y)  &  =\frac{1}{a_{0}}\sum_{n}E_{\times}\left(  y-nb_{0}\right)
\label{CrossEnColl}\\
&  \approx-\frac{\Phi_{0}^{2}}{4\pi^{2}\gamma^{2}sa_{0}\ln(\gamma s/\lambda
)}\sum_{m=-\infty}^{\infty}A_{\times}\left(  \frac{y-mb_{0}}{\gamma s}\right)
\nonumber
\end{align}
where $b_{0}=\sqrt{3}a_{0}/2$ is the distance between the PV rows.
This expression has a logarithmic accuracy, i.e., we neglected a
weak $y$-dependence under the logarithm and replaced the
dimensionless function under the logarithm by its typical value.
Using Fourier transform of $E_{\times}(y) $, $\tilde{E}
_{\times}(p)=\int dy\exp(-ipy)E_{\times}(y)$, we also represent
$E_{Jl}(y)$ as
\[
E_{Jl}(y)=n_{v}\sum_{s=-\infty}^{\infty}\tilde{E}_{\times}\left(  \frac{2\pi
s}{b_{0}}\right)  \exp\left(  i\frac{2\pi m}{b_{0}}y\right)
\]
In the case $b_{0}\ll\gamma s$ we can keep only $m=0,\pm1$ terms
in this sum,
\[
E_{Jl}(y)\approx E_{Jl}(0)+2n_{v}\tilde{E}_{\times}\left(
\frac{2\pi}{b_{0} }\right)  \cos\left(  \frac{2\pi
y}{b_{0}}\right)
\]
Using interpolation formula (\ref{AcrosInt}) for $A_{\times}\left(
\tilde {y}\right) $,
we obtain the asymptotics for $b_0<\gamma s$
\begin{align}
E_{Jl}(y)-E_{Jl}(0)&\approx-0.35\frac{n_{v}\Phi_{0}^{2}}{\gamma\ln(\gamma
s/\lambda)}\sqrt{\frac{\gamma s}{b_{0}}} \label{LatEnAsymp}\\
&\times \cos\left(  \frac{2\pi y}{b_{0} }\right)  \exp\left(
-5.82\frac{\gamma s}{b_{0}}\right)\nonumber
\end{align}

To compute the field dependence of the critical current, we
represent the force acting on JV from the pancake lattice in the
scaling form,
\[
F_{Jpl}(y)\approx\frac{\sqrt{3}s\Phi_{0}^{2}}{(4\pi)^{2}(\gamma
s)^{4} \ln(\lambda/s)}\mathcal{F}\left(  \frac{y}{\gamma
s},\frac{b_{0}}{\gamma s}\right),
\]
with
\[
\mathcal{F}(\tilde{y},\tilde{b}_{0})\equiv\frac{2}{\tilde{b}_{0}}\frac
{d}{d\tilde{y}}\sum_{j}A_{\times}(\tilde{y}-j\tilde{b}_{0}).
\]
The critical pinning current is given by
\begin{equation}
j_{Jpl}=j_{J}\frac{\sqrt{3}\lambda^{2}}{2(\gamma s)^{2}\ln(\lambda
/s)}\mathcal{F}_{c}(\tilde{b}_{0}) \label{JplB}
\end{equation}
with
\[
\mathcal{F}_{c}(\tilde{b}_{0})=\max_{\tilde{y}}[\mathcal{F}(\tilde{y}
,\tilde{b}_{0})]
\]
\begin{figure}
[ptb]
\begin{center}
\includegraphics[width=3.4in ] {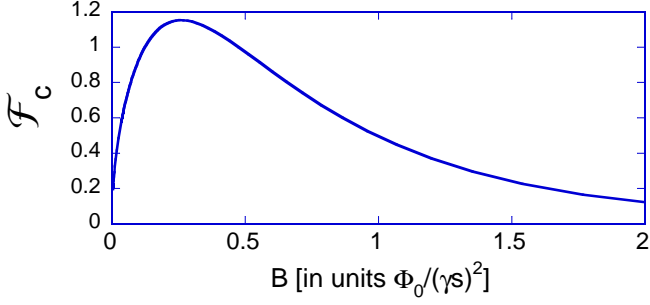}
\caption{The field dependence of the dimensionless pinning force
of JV by dilute pancake vortex crystal}
\end{center}
\end{figure}
Numerically calculated dependence of $\mathcal{F}_{c}$ versus
reduced field $(\gamma
s)^{2}B/\Phi_{0}\equiv\sqrt{3}/2\tilde{b}_{0}^{2}$. Maximum
$\mathcal{F}_{c\max}\approx1.15$ is achieved at
$B\approx0.26\Phi_{0}/(\gamma s)^{2}$ ($b_{0}=2.1\gamma s$).
Therefore the maximum pinning current can be estimated as
\begin{equation}
j_{Jp\max}\approx j_{J}\frac{\lambda^{2}}{(\gamma
s)^{2}\ln(\lambda/s)}
\end{equation}
For typical parameters of BSCCO this current is only 5-10 times
smaller than the maximum Josephson current, i.e., it is actually
rather large.

An exponential decay of the pinning energy (\ref{LatEnAsymp}) and
force holds until the row separation $b_0$ reaches $\lambda$. At
larger fields one have to take into account shrinking of the
vortex core. Because the number of pancake rows in the core is
almost constant at high field, the exponential decay will saturate
at a finite value $\sim \exp(-C\gamma s/\lambda)$.

\section{\label{Sec:JVVisc}Pancake vortices and viscosity of Josephson vortex}

A sufficiently large $c$ axis transport current will drive JVs.
Two dynamic regimes are possible, depending on the relation
between the JV-pancake interaction and disorder-induced pancake
pinning. Moving JVs either can drag the pancake lattice or they
can move through the static pancake lattice.  Slow dragging of
pancake stacks by JVs at small $c$ axis fields have been observed
experimentally.\cite{GrigNat01} As the JV-pancake interaction
force decays exponentially at high $c$ axis fields in the case
$\lambda < \gamma s$, one can expect that JVs will always move
through the static pancake lattice at sufficiently high fields.

\subsection{Dragging pancake lattice by Josephson vortices}

When moving JVs drag the pancake lattice, one can obtain simple
universal formulas for the JV viscosity coefficient and JV
flux-flow resistivity. The effective viscosity coefficient per
single JV is connected by a simple relation with the viscosity
coefficient of pancake stack per unit length, $\eta_{v}$,
\[
B_{x}\eta_{JV}=B_{z}\eta_{v}.
\]
Therefore, the JV flux-flow resistivity,
$\rho_{ff}^{c}(\mathbf{B})=\Phi_{0}B_{x}/(c^{2}\eta_{JV})$, is
given by
\begin{equation}
\rho_{ff}^{c}(\mathbf{B})=\frac{\Phi_{0}B_{x}^{2}}{c^{2}\eta_{v}B_{z}}.
\label{rhoff-drag}
\end{equation}
As a consequence, we also obtain a simple relation between
$\rho_{ff}^{c}(\mathbf{B})$ and $\rho_{ff}^{ab}(B_{z})$
\[
\rho_{ff}^{c}(\mathbf{B})=\rho_{ff}^{ab}(B_{z})\frac{B_{x}^{2}}{B_{z}^{2}}.
\]

\subsection{Josephson vortex moving through pancake lattice}

Consider slow JV motion through the static pancake lattice.  JV
motion along the $y$ direction with velocity $V_{y}$ induces
traveling pancake displacement field $u_{n}(y-V_{y}t)$.
For slow motion, $u_{n}(y)$ is just
the static displacement field around JV. Contribution to the
energy dissipation caused by these displacements is given by
\begin{align*}
W  & \approx\eta_{p}n_{v}\sum_{n}\int
d\mathbf{r}\dot{u}_{n}^{2}\\
&  =\eta_{p}\left[  n_{v}\sum_{n}\int d\mathbf{r}\left(  \mathbf{\nabla}
_{y}u_{n}\right)  ^{2}\right]  V_{y}^{2}
\end{align*}
where $\eta_{p}$ is the pancake viscosity coefficient. Therefore
the JV viscosity per unit length is given by
\begin{equation}
\eta_{JV}=\eta_{p}n_{v}\sum_{n}\int d\mathbf{r}\left(
\mathbf{\nabla} _{y}u_{n}\right)  ^{2} \label{viscJV}
\end{equation}
Using relation between the vortex phase and displacement field
$\nabla_{y} \phi_{vn}=2\pi n_{v}u_{n}$ we obtain an estimate
\[
n_{v}\sum_{n}\int dy\left(  \mathbf{\nabla}_{y}u_{n}\right)
^{2}\approx \frac{2.7\lambda^{3}\sqrt{n_{v}}}{(\gamma
s)^{3}}\left(  \frac{\mathcal{C} }{\ln\left( r_{\rm
cut}/r_{w}\right) }\right)  ^{3/2}
\]
with $\mathcal{C}\equiv \langle \cos(\phi_{n+1}-\phi_n) \rangle <1
$, and derive an approximate formula for the JV viscosity
coefficient
\begin{equation}
\eta_{JV}\approx\eta_{v}\frac{s}{a}\frac{2.7\lambda^{3}}{(\gamma
s)^{3}}\left( \frac{\mathcal{C}}{\ln\frac{r_{cut}}{r_{w}}}\right)
^{3/2} \label{viscJVestim}
\end{equation}
with $\eta_{v}\equiv\eta_{p}/s$ is the viscosity of pancake stack
per unit length. Because the JV viscosity $\eta_{JV}$ is
proportional to the pancake-stack viscosity $\eta_{v}$, there is a
relation between the flux-flow resistivity of JVs
($\rho_{ff}^{c}$) and the flux-flow resistivity of pancake
vortices ($\rho_{ff}^{ab}$)
\begin{equation}
\frac{\rho_{ff}^{c}}{B_{x}}\approx
\frac{\rho_{ff}^{ab}}{B_{z}}\frac{a}{s} \frac{0.37(\gamma
s)^{3}}{\lambda^{3}}\left( \frac{\ln(r_{cut}/r_{w})
}{\mathcal{C}}\right)  ^{3/2} \label{rhoc-rhoab}
\end{equation}
If we use the Bardeen-Stephen formula for the in-plane flux-flow
resistivity, $\rho_{ff}^{ab}\approx\rho_{ab}B_{z}/H_{c2}$, and an
estimate for the $c$ axis flux-flow resistivity at $B_z=0$,
$\rho_{ff}^{c0}\approx(16\gamma^{3}s^{2}B_{x}/\Phi_{0})\rho_{ab}$,\cite{KoshPRB00}
we can also obtain relation between $\rho_{ff}^{c}(B_{z})$ and
$\rho_{ff}^{c0}$
\begin{equation}
\rho_{ff}^{c}\approx 0.15\rho_{ff}^{c0}\frac{\xi^{2}
a}{\lambda^{3}}\left(
\frac{\ln(r_{cut}/r_{w})}{\mathcal{C}}\right) ^{3/2}
\label{rhoc-rhoc}
\end{equation}
From this estimate we can see that at $a\sim\lambda$ the flux-flow
resistivity for JVs slowly moving through the pancake lattice is
about four order of magnitude (factor $(\xi/\lambda)^{2}$) smaller
than the flux-flow resistivity of free JVs. We see that even
though the critical force becomes exponentially small at high
fields, pancakes still very strongly hinder mobility of JVs.

\section{Acknowledgements}

I would like to thank V.\ Vlasko-Vlasov, T.\ Tamegai, A.\
Grigorenko, and S.\ Bending for fruitful discussions. I am very
grateful to M.\ Dodgson for reading the manuscript and useful
suggestions. This work was supported by the U.\ S.\ DOE, Office of
Science, under contract \# W-31-109-ENG-38.

\appendix
\section{\label{App-MagIntRows}Magnetic interaction between pancake rows}

In this appendix we derive several useful representation for the
magnetic interaction between pancake rows in different layers. The
interaction energy between two pancakes in Fourier and real-space
is given by Eq.\ (\ref{MagInter}). We will use also this
interaction in the mixed representation
\[
U_{M}(\mathbf{k}_{\perp},n)\approx\frac{4\pi^{2}J\delta_{n}}{k_{\perp}^{2}}
-\frac{2\pi^{2}sJ}{\lambda^{2}k_{\perp}^{2}}\frac{\exp\left(
-ns\sqrt{\lambda^{-2}+k_{\perp}^{2}}\right)
}{\sqrt{\lambda^{-2}+k_{\perp }^{2}}}
\]
The interaction energy between the pancake rows per unit length is
given by
\begin{widetext}
\begin{align*}
U_{Mr}(x,y,n)  &  \equiv\frac{1}{a}\sum_{m}U_{M}(x-ma,y,n)\\
&
=\frac{1}{a^{2}}\sum_{l}\int\frac{dk_{y}}{2\pi}U_{M}(k_{x}=\frac{2\pi
l} {a},k_{y},n)\exp\left(  i\frac{2\pi l}{a}x+ik_{y}y\right)
\end{align*}
From this equation we obtain the following integral representation
for $U_{Mr}(x,y,n)$
\begin{align*}
U_{Mr}(x,y,n)  &  =\frac{s\varepsilon_{0}}{a}\left[-\left(
\delta_{n} -\frac{s}{2\lambda}\exp\left(
-\frac{s|n|}{\lambda}\right) \right) \ln\left[  1-2\cos\frac{2\pi
x}{a}\exp\left( -\frac{2\pi|y|}{a}\right)
+\exp\left(  -\frac{4\pi|y|}{a}\right)  \right]\right . \\
&  \left.+\frac{s}{a}\sum_{l}\int du\frac{\exp\left(  i\frac{2\pi
l}{a} x+i\sqrt{\lambda^{-2}+\left(  \frac{2\pi l}{a}\right)
^{2}}usn-\sqrt {\lambda^{-2}+\left(  \frac{2\pi l}{a}\right)
^{2}}\sqrt{1+u^{2}}|y|\right) }{\left(  1+\left(  1+\left(
\frac{2\pi\lambda l}{a}\right)  ^{2}\right) u^{2}\right)
\sqrt{1+u^{2}}}\right]
\end{align*}
This representation can be used to derive large-$y$ asymptotics of
$U_{Mr}(x,y,n)$,
\begin{align*}
U_{Mr}(x,y,n)  &  \approx\frac{\pi J}{a}\left[\left(
\delta_{n}-\frac {s}{2\lambda}\exp\left(
-\frac{s|n|}{\lambda}\right)  \right) 2\cos
\frac{2\pi x}{a}\exp\left(  -\frac{2\pi|y|}{a}\right)\right. \\
&  \left.+\frac{\sqrt{2\pi}s}{a}\frac{\exp\left(
i\frac{2\pi}{a}x-\sqrt {\lambda^{-2}+\left(  \frac{2\pi}{a}\right)
^{2}}|y|-\frac{\sqrt{\lambda ^{-2}+\left(  \frac{2\pi}{a}\right)
^{2}}\left(  sn\right)  ^{2}} {2|y|}\right)  }{\left(
\lambda^{-2}+\left(  \frac{2\pi}{a}\right) ^{2}\right)
^{1/4}\sqrt{|y|}}\right]
\end{align*}
at$\sqrt{\lambda^{-2}+\left(  2\pi /a\right) ^{2}}|y|\gg1$.
Therefore the interaction between the pancake rows decays
exponentially at $y,sn >\lambda,a$

Interaction of pancake row with stack of rows is given by
\begin{align*}
U_{Ms}(x,y)  &  =2\sum_{n=1}^{\infty}U_{Mr}(x,y,n)\\
&  =\frac{\pi J}{a}\ln\left[  1-2\cos(\frac{2\pi x}{a})\exp\left(  -\frac{2\pi
y}{a}\right)  +\exp\left(  -\frac{4\pi y}{a}\right)  \right] \\
&  +\frac{2\pi^{2}J}{a}\sum_{l}\frac{\exp\left(  i\frac{2\pi l}{a}
x-\sqrt{\lambda^{-2}+\left(  \frac{2\pi l}{a}\right)  ^{2}}y\right)  }
{\sqrt{(a/\lambda)^{2}+\left(  2\pi l\right)  ^{2}}}
\end{align*}
In particular, the potential created by pancakes belonging to the
same row stack (``cage potential'') is given by

\[
U_{\mathrm{cage}}\equiv
U_{Ms}(x,0)=-\frac{4\pi^{2}J}{a}\sum_{l=1}^{\infty }\cos\left(
\frac{2\pi l}{a}x\right)  \left(  \frac{1}{2\pi l}-\frac{1}
{\sqrt{(a/\lambda)^{2}+\left(  2\pi l\right)  ^{2}}}\right)
\]
\end{widetext}

\section{\label{AppEJosc} Local contribution to Josephson energy
due to mismatch of pancake rows in neighboring layers}

We consider two pancake rows in neighboring layers with period
$a\ll\gamma s $ shifted at distance $v$ with respect to each other
in the direction of row ($x$ axis). In zero order with respect to
the Josephson coupling these rows produce the phase mismatch
$\varphi_{v}(x,y)$ between the layers,
\[
\varphi_{v}(x,y)\!=\!\sum_{m}\left(\!
\arctan\frac{x\!-\!m\!+\!v/2}{y}-\arctan
\frac{x\!-\!m\!-\!v/2}{y}\right) .
\]
We measure all distances in units of the lattice constant $a$. The
$x$-averaged phase mismatch is given by
\[
\overline{\varphi}_{v}(y)=\pi v\,\mathrm{sgn}(y).
\]
The total phase approaches $\overline{\varphi}_{v}(y)$ at $y\gtrsim1/2\pi$.

Separating $\varphi_{v}(x,y)$ from the total phase, we can represent the
Josephson energy as
\begin{widetext}
\begin{align*}
\mathcal{F}_{J}  &  =E_{J}a^{2}\int d^{2}\mathbf{r}\left[  1-\cos\left(
\varphi_{v}(x,y)+\varphi(x,y)\right)  \right] \\
&  =E_{J}a^{2}\int d^{2}\mathbf{r}\left[  1-\cos\left(
\overline{\varphi} _{v}(y)+\varphi(x,y)\right)  \right]
+L_{x}\mathcal{E}_{Josc}(v,\varphi)
\end{align*}
where $\varphi(x,y)$ is the smooth external phase and the local Josephson
energy $\mathcal{E}_{Josc}(v,\varphi)$ per unit length is defined as
\begin{align}
\mathcal{E}_{Josc}(v,\varphi)  &  =-\frac{E_{J}a^{2}}{L_{x}}\int dxdy\left[
\cos\left(  \varphi+\varphi_{v}(x,y)\right)  -\cos\left(  \varphi
+\overline{\varphi}_{v}(y)\right)  \right] \label{OscJosEn}\\
&  =2E_{J}a\cos\left(  \varphi\right)  \mathcal{J}\nonumber
\end{align}
where
\begin{equation}
\mathcal{J}(v)\equiv\int_{-1/2}^{1/2}dx\int_{0}^{\infty}dy\left[  \cos\left(
\overline{\varphi}_{v}(y)\right)  -\cos\left(  \varphi_{v}(x,y)\right)
\right]  . \label{IntEJosc}
\end{equation}
In $\mathcal{E}_{Josc}(v,\varphi)$ we can neglect weak coordinate
dependence of the external phase and replace it by a constant
$\varphi$. At $|v|<1/2$ the ground state for fixed $v$ corresponds
to $\varphi=0$, while at $1/2<|v|<1$ the ground state corresponds
to $\varphi=\pi$. $\mathcal{E}_{Josc}(v,\varphi)$ has a symmetry
property $\mathcal{E}_{Josc}(1-v,\varphi)=-\mathcal{E}
_{Josc}(v,\varphi)$. The integral over $y$ in $\mathcal{J}(v)$ is
converges at $y\lesssim1/2\pi$. This allows us to consider a
single row separately from other rows and neglect the coordinate
dependence of the ``external phase'' $\varphi$.

Using the complex variable $z=x+iy$ one can derive a useful expression for
$\varphi_{v}(x,y)$:
\[
\varphi_{v}(z)=\operatorname{Im}\left(  \ln\prod_{m}\frac{z-m-v/2}
{z-m+v/2}\right)  =\operatorname{Im}\left(  \ln\frac{\sin\left(
\pi(z-v/2)\right)  }{\sin\left(  \pi(z+v/2)\right)  }\right)  .
\]
Going back to the $(x,y)$ representation, we obtain
\[
\varphi_{v}(x,y)=\arctan\frac{\tan\left(  \pi(x+v/2)\right)  }{\tanh\pi
y}-\arctan\frac{\tan\left(  \pi(x-v/2)\right)  }{\tanh\pi y}
\]
and
\[
\cos\varphi_{v}(x,y)=\frac{\cosh2\pi y\cos\pi v-\cos2\pi x}{\sqrt{(\cosh2\pi
y-\cos2\pi x\cos\pi v)^{2}-(\sin2\pi x\sin\pi v)^{2}}}.
\]
Integral (\ref{IntEJosc}) can now be represented as
\[
\mathcal{J}(v)=\int_{-1/2}^{1/2}dx\int_{0}^{\infty}dy\left[  \cos\left(  \pi
v\right)  -\frac{\cosh2\pi y\cos\pi v-\cos2\pi x}{\sqrt{(\cosh2\pi y-\cos2\pi
x\cos\pi v)^{2}-(\sin2\pi x\sin\pi v)^{2}}}\right]
\]
We obtain an approximate analytical result at small $v$, $v\ll1$. Simple
expansion with respect to $v$ produces logarithmically diverging integral. To
handle this problem we introduce the intermediate scale $y_{0}$, $v\ll
y_{0}\ll1$, and split integral $\mathcal{J}$ into contribution coming from
$y>y_{0}$ ($\mathcal{J}_{>}$) and $y<y_{0}$ ($\mathcal{J}_{<}$). At $y>y_{0}$
we use small-$v$ expansion and obtain
\begin{align*}
\mathcal{J}_{>}  &  =\left(  \pi v\right)  ^{2}\int_{y_{0}}^{\infty}dy\int
_{0}^{1/2}dx\left(  -1+\frac{\sinh^{2}2\pi y}{\left(  \cosh2\pi y-\cos2\pi
x\right)  ^{2}}\right) \\
&  =\frac{\left(  \pi v\right)  ^{2}}{2}\int_{y_{0}}^{\infty}dy\frac
{\exp(-2\pi y)}{\sinh2\pi y}\\
&  \approx\frac{\pi v^{2}}{4}\ln\frac{1}{4\pi y_{0}}.
\end{align*}
In region $y<y_{0}$ we can expand all trigonometric functions and obtain the
integral
\begin{align*}
\mathcal{J}_{<}  &  \approx2\int_{0}^{y_{0}}dy\int_{0}^{1/2}dx\left[
1-\frac{4(x^{2}+y^{2})-v^{2}}{\sqrt{(4\left(  x^{2}+y^{2}\right)  +v^{2}
)^{2}-(4xv)^{2}}}\right] \\
&
=\frac{v^{2}}{2}\int_{0}^{2y_{0}/v}d\tilde{y}\int_{0}^{1/v}d\tilde
{x}\left[
1-\frac{\tilde{x}^{2}+\tilde{y}^{2}-1}{\sqrt{(\tilde{x}^{2}
+\tilde{y}^{2}-1)^{2}+4\tilde{y}^{2}}}\right]
\end{align*}
\end{widetext}
with $\tilde{y}=2y/v$ and $\tilde{x}=2x/v$. Because we only
interested in the main logarithmic term, we can extend integration
over $\tilde{x}$ up to $\infty$. The obtained integral can be
evaluated as
\[
\mathcal{J}_{<}\approx\frac{\pi}{4}v^{2}\left(  \ln\frac{y_{0}}{v}
+1.58\right)
\]
Adding $\mathcal{J}_{>}$ and $\mathcal{J}_{<}$, we obtain
\begin{equation}
\mathcal{J}(v)\approx\frac{\pi}{4}v^{2}\left(  \ln\frac{1}{v}-0.95\right)
\label{J-small-v}
\end{equation}
and
\begin{equation}
E_{Josc}(v,\varphi)\approx\frac{\pi}{2}E_{J}a\cos\left(  \varphi\right)
v^{2}\left(  \ln\frac{1}{v}-0.95\right)  \label{EJosc-small-v}
\end{equation}
\begin{figure*}
[ptb]
\begin{center}
\includegraphics[width=3.4in]{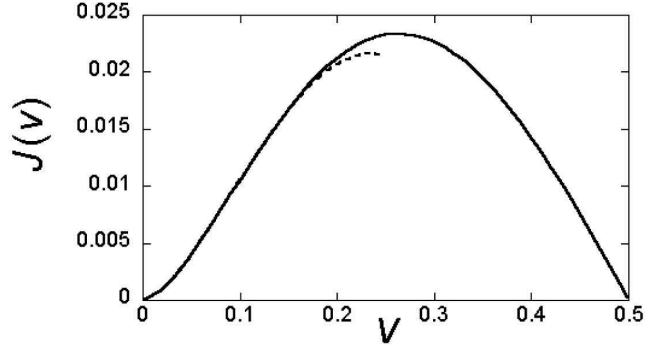}
\caption{Dimensionless function $\mathcal{J}(v)$ which
determines the local
Josephson energy. Dashed line is small-$v$ asymptotics (\ref{J-small-v}).}
\label{Fig-EJosc}
\end{center}
\end{figure*}
We also calculated function $\mathcal{J}(v)$ numerically for the whole range
$0<v<1/2 $. The result is shown in Fig.\ \ref{Fig-EJosc} and is described by
approximate interpolation formula
\begin{align}
\mathcal{J}(v)&\approx \frac{1-\cos\pi v}{4\pi}\label{J-interp}\\
&\times \left( \ln\frac{1}{1-\cos\pi v}-0.379\cos\pi
v+0.076\cos^{2}\pi v\right)\nonumber
\end{align}
This result was used in numerical calculations of the JV core
structure.

\section{\label{App:ContrLocalTerm} Contribution to JV energy coming from local Josephson
term at \lowercase{$\gamma\gg\lambda/s$}}

In the limit of very weak coupling the correction to reduced JV energy
(\ref{RedEn}) coming from the local Josephson energy is given by
\begin{widetext}
\begin{equation}
\delta\mathcal{E}_{JV}\approx\frac{\pi\sqrt{3}}{4\sqrt{1+B_{z}/B_{\lambda}}
}\sum_{n=-\infty}^{\infty}\int_{-\infty}^{\infty}d\tilde{y}(v_{n+1}-v_{n}
)^{2}\ln\frac{0.39}{\left|  v_{n+1}-v_{n}\right|
}\cos(\phi_{n+1}-\phi_{n})
\label{dEJVosc}
\end{equation}
with $\tilde{y}=y/\lambda_{J}=\sqrt{1+B_{z}/B_{\lambda}}y/\gamma s$. We will
focus only on the regime $B_{z}\gg B_{\lambda}$, where this correction can be
noticeable. In this regime reduced row displacements are connected with phase
gradient by relation
\[
v_{n}(y)\approx\frac{b}{2\pi\lambda_{J}}\tilde{\nabla}_{y}\phi_{n}
\]
and the correction reduces to
\begin{equation}
\delta\mathcal{E}_{JV}\approx\frac{\sqrt{B_{z}/B_{\lambda}}}{8\pi
n_{v}\left( \gamma s\right)  ^{2}}\sum_{n}\int d\tilde{y}\left(
\tilde{\nabla}_{y} \phi_{n+1}-\tilde{\nabla}_{y}\phi_{n}\right)
^{2}\ln\left[  \frac {2.83\,\gamma
s\sqrt{B_{\lambda}/B_{z}}}{a\left|  \tilde{\nabla}_{y}\phi
_{n+1}-\tilde{\nabla}_{y}\phi_{n}\right|  }\right]
\cos(\phi_{n+1}-\phi_{n}).
\end{equation}
Using numerical estimates
\begin{align*}
\sum_{n=-\infty}^{\infty}\int_{-\infty}^{\infty}d\tilde{y}\cos\left(
\phi_{n+1}-\phi_{n}\right)  \left(  \tilde{\nabla}_{y}\phi_{n+1}-\tilde
{\nabla}_{y}\phi_{n}\right)  ^{2}  &  \approx-2.4\\
\sum_{n=-\infty}^{\infty}\int_{-\infty}^{\infty}d\tilde{y}\cos\left(
\phi_{n+1}-\phi_{n}\right)  \left(
\tilde{\nabla}_{y}\phi_{n+1}-\tilde {\nabla}_{y}\phi_{n}\right)
^{2}\ln\frac{1}{\left|  \tilde{\nabla}_{y}
\phi_{n+1}-\tilde{\nabla}_{y}\phi_{n}\right|  }  &  \approx4.7
\end{align*}
obtained with the JV\ phase $\phi_{n}(\tilde{y})$ (\ref{JVPhase}), we obtain
\begin{equation}
\delta\mathcal{E}_{JV}\approx-\pi\sqrt{\frac{B_{\lambda}}{B_{z}}}
\frac{0.38\lambda^{2}}{\left(  \gamma s\right)  ^{2}\ln\left(
a/r_{w}\right) }\ln\left[  \frac{0.1\;\gamma
s}{\lambda}\sqrt{\ln\frac{a}{r_{w}}}\right]  .
\end{equation}
As we can see, the correction is smaller than the reduced JV\
energy at $B_{z}\gg B_{\lambda}$,
$\mathcal{E}_{JV}\approx\pi\sqrt{B_{\lambda}/B_{z}}\ln(a/s)$, by
the factor $\sim \lambda^{2}/(\gamma s)^{2}$.
\end{widetext}


\begin{thebibliography}{99}

\bibitem{pancakes}A.\ I.\ Buzdin and D.\ Feinberg, J. Phys. (Paris)
\textbf{51}, 1971 (1990); S.\ N.\ Artemenko and A.\ N.\ Kruglov,
Phys.\ Lett.\ A \textbf{143} 485 (1990); J.\ R.\ Clem, Phys.\ Rev.\ B
\textbf{43}, 7837 (1991).


\bibitem{BulJosLat}L.\ N.\ Bulaevskii, Zh.\ Eksp.\ Teor.\ Fiz.\ \textbf{64},
2241 (1973) [Sov.\ Phys.\ JETP \textbf{37}, 1133 (1973)].

\bibitem{ClemJosLat}J.\ R.\ Clem and M.\ W.\ Coffey, Phys.\ Rev.\ B,
\textbf{42}, 6209 (1990); J.\ R.\ Clem, M.\ W.\ Coffey, and Z.\ Hao,
Phys.\ Rev.\ B \textbf{44}, 2732 (1991).

\bibitem{Kinkwalls}A.\ E.\ Koshelev, Phys.\ Rev.\ B \textbf{48}, 1180 (1993).

\bibitem{BLK}L.\ N.\ Bulaevskii, M.\ Ledvij, and V.\ G.\ Kogan,
Phys.\ Rev.\ B \textbf{46}, 366 (1992); M.\ Benkraouda and M.\ Ledvij,
Phys.\ Rev.\ B \textbf{51}, 6123 (1995).

\bibitem{CrossLatLet}A.\ E.\ Koshelev, Phys.\ Rev.\ Lett. \textbf{83}, 187 (1999).

\bibitem{Feinberg93}D.\ Feinberg and A.\ M.\ Ettouhami, Int.\ J.\ Mod.\ Phys.
B \textbf{7}, 2085 (1993).

\bibitem{chains}A.\ I.\ Buzdin and A.\ Yu.\ Simonov, JETP Lett.\ \textbf{51},
191 (1990); A.\ M.\ Grishin, A.\ Y.\ Martynovich, and %
S.\ V.\ Yampolskii, Sov.\ Phys.\ JETP \textbf{70}, 1089 (1990);
B.\ I.\ Ivlev and N.\ B.\ Kopnin, Phys.\ Rev.\ B \textbf{44}, 2747
(1991); W.\ A.\ M.\ Morgado, M.\ M.\ Doria, and G.\ Carneiro,
Physica C, \textbf{349}, 196 (2001).

\bibitem{coex-lat}L.\ L.\ Daemen, L.\ J.\ Campbell,
A.\ Yu.\ Simonov, and V.\ G.\ Kogan Phys.\ Rev.\ Lett.\ \textbf{70}, 2948
(1993); A.\ Sudb{\o }, E.\ H.\ Brandt, and D.\ A.\ Huse
Phys.\ Rev.\ Lett.\ \textbf{71}, 1451 (1993); E.\ Sardella, Physica C
\textbf{257}, 231 (1997).


\bibitem{BvsH}Direction of magnetic induction $\mathbf{B}$ almost never
coincides with direction of external magnetic field
$\mathbf{H}_{ext}$. The relation between $\mathbf{B}$ and
$\mathbf{H}_{ext}$ depends on the shape of sample and constitutes
a separate problem.

\bibitem{Savel01}S.\ E.\ Savel'ev, J.\ Mirkovic, K.\ Kadowaki, Phys.\ Rev.\ B
\textbf{64}, 094521 (2001).

\bibitem{Schmidt}B.\ Schmidt,M.\ Konczykowski, N.\ Morozov, and E.\ Zeldov,
Phys.\ Rev.\ B \textbf{55}, R8705 (1997).

\bibitem{OoiPRL99}S.\ Ooi, T.Shibauchi, N.Okuda, and T.Tamegai,
Phys.\ Rev.\ Lett.\ \textbf{82}, 4308 (1999).


\bibitem{Huse}D.\ A.\ Huse, Phys.\ Rev.\ B \textbf{46},8621 (1992).

\bibitem{Bolle91}C.\ A.\ Bolle,  P.\ L.\ Gammel, D.\ G.\ Grier, C.\ A.\ Murray,
D.\ J.\ Bishop, D.\ B.\ Mitzi, and A.\ Kapitulnik , Phys.\ Rev.\
Lett.\ \textbf{66}, 112 (1991).

\bibitem{Grig95}I.\ V.\ Grigorieva, J.\ W.\ Steeds, G.\
Balakrishnan, and D.\ M.\ Paul, Phys.\ Rev.\ B \textbf{51}, 3765
(1995).

\bibitem{GrigNat01}  A.\ Grigorenko, S.\ Bending, T.\ Tamegai, S.\ Ooi, and
M.\ Henini, Nature {\bf 414}, 728 (2001).

\bibitem{MatsudaSci02}T.\ Matsuda, O.\ Kamimura, H.\ Kasai, K.\ Harada,
T.\ Yoshida, T.\ Akashi, A.\ Tonomura, Y.\ Nakayama, J.\
Shimoyama, K.\ Kishio, T.\ Hanaguri, and K.\ Kitazawa, Science,
\textbf{294}, 2136(2001).

\bibitem{VlaskoPRB02}V.\ K.\ Vlasko-Vlasov, A.\ E.\ Koshelev,  U.\ Welp,%
G.\ W.\ Crabtree, and K.\ Kadowaki, Phys.\ Rev.\ B \textbf{66},
014523 (2002).

\bibitem{TokunagaPRB02}M.\ Tokunaga, M.\ Kobayashi, Y.\ Tokunaga, and
T.\ Tamegai, Phys.\ Rev.\ B \textbf{66}, 060507 (2002).

\bibitem{DodgsonPRB02} M.\ J.\ W.\ Dodgson, Phys.\ Rev.\ B \textbf{66}, 014509
(2002).

\bibitem{BuzdinPRL02}  A.\ Buzdin and I.\ Baladi\'{e}, Phys.\ Rev.\ Lett.\
\textbf{88}, 147002 (2002).

\bibitem{KonczPhysC00}M.\ Konczykowski, C.\ J.\ van der Beek, M.\ V.\ Indenbom,%
E.\ Zeldov, Physica C, \textbf{341}, 1213 (2000).

\bibitem{MirkovPRL01} J.\ Mirkovic, S.\ E.\ Savelev, E.\ Sugahara, and K.\ Kadowaki,
Phys.\ Rev.\ Lett.\ \textbf{86}, 886 (2001).

\bibitem{OoiPRB01}S.\ Ooi, T.\ Shibauchi, K.\ Itaka, N.\ Okuda, and %
T.\ Tamegai, Phys.\ Rev.\ B \textbf{63}, 020501(R) (2001)

\bibitem{TokunagaPRB02A}M.\ Tokunaga, M.\ Kishi, N.\ Kameda, K.\ Itaka, and %
T.\ Tamegai, Phys.\ Rev B \textbf{66}, 220501(R) (2002).

\bibitem{TiltSCHA}The more quantitative expression for
$U_{44}(\mathbf{k})$, which accounts for thermal softening, can be
obtained using the self-consistent harmonic approximation. At high
fields and $k_{z}>r_{w}^{-1}$ it gives
$U_{44}(\mathbf{k})\approx\frac
{n_{v}\varepsilon_{0}}{2\lambda^{2}}\ln\left(
0.5+\frac{0.13a^{2}}{r_{w}^{2} }\right)$, see A.\ E.\ Koshelev and
L.\ N.\ Bulaevskii Physica C, \textbf{341-348}, 1503 (2000).

\bibitem{PhaseDistNote}The formula (B12) for the JV phase in
Ref.\ \onlinecite{Kinkwalls} for phase distribution $\phi_{n}$ was written
incorrectly. Eq.\ (\ref{JVPhase}) gives correct phase distribution, which was
actually used in all numerical calculations and is plotted in Fig.\ 8 of
Ref.\ \onlinecite{Kinkwalls}.

\bibitem{KoshKes93}A.\ E.\ Koshelev and P.\ H.\ Kes, Phys.\ Rev.\ B
\textbf{48}, 6539 (1993).

\bibitem{Horovitz}T.\ R.\ Goldin and B.\ Horovitz, Phys.\ Rev.\ B
\textbf{58}, 9524 (1998).

\bibitem{noteDodgsonPRL99} The identical factor
$1+B_z/B_{\lambda}$ appears also in the renormalization of defect
energy by the pancake vortex crystal and have precisely the same
physical origin, see M.\ J.\ W.\ Dodgson, V.\ B.\ Geshkenbein, and
G.\ Blatter, Phys.\ Rev.\ Lett.\ \textbf{83}, 5358 (1999).

\bibitem{LatPhysC91}Yu.\ Latyshev and A.\ Volkov, Physica C, \textbf{182}, 47
(1991).

\bibitem{EnriquezPRB96}H.\ Enriquez, N.\ Bontemps, P.\ Fournier,
A.\ Kapitulnik, A.\ Maignan, and A.\ Ruyter, Phys.\ Rev.\ B,
\textbf{53}, R14757 (1996)

\bibitem{HechtfischerPRB97}G.\ Hechtfischer, R.\ Kleiner, K.\ Schlenga,
W.\ Walkenhorst, P.\ M\"{u}ller, and H.\ L.\ Johnson, Phys.\ Rev.\
B \textbf{55}, 14638 (1997).
%

\bibitem{KoshPRB00}A.\ E.\ Koshelev, Phys.\ Rev.\ B \textbf{62}, R3616 (2000).



\end{thebibliography}
\end{document}